\documentclass[conference]{IEEEtran}

\IEEEoverridecommandlockouts
\usepackage{amsmath,amssymb,amsfonts}
\usepackage{mathabx}
\usepackage{amsthm}
\usepackage{nicefrac
}
\newtheorem{theorem}{Theorem}

\usepackage{algorithmic}
\usepackage{url}
\usepackage{graphicx}
\usepackage{textcomp}
\usepackage{xcolor}
\def\BibTeX{{\rm B\kern-.05em{\sc i\kern-.025em b}\kern-.08em
    T\kern-.1667em\lower.7ex\hbox{E}\kern-.125emX}}

\usepackage{subfig}
\usepackage[nolist,nohyperlinks]{acronym}
\usepackage{siunitx}
\usepackage{cite}

\usepackage{microtype}
\usepackage{nicefrac}
\usepackage{multirow}
\usepackage{rotating}
\usepackage{caption}  
\usepackage[table]{xcolor}
\usepackage{microtype}
\makeatletter
\newcommand\footnoteref[1]{\protected@xdef\@thefnmark{\ref{#1}}\@footnotemark}
\makeatother

\DeclareMathOperator*{\argmin}{arg\,min}	

\acrodef{0-D}[0-D]{zero-dimensional}
\acrodef{1-D}[1-D]{one-dimensional}
\acrodef{3-D}[3-D]{three-dimensional}
\acrodef{MC}[MC]{synthetic molecular communication}
\acrodef{EM}[EM]{electromagnetic wave}
\acrodef{DOI}[DOI]{digital object identifier}
\acrodef{CGF}[CGF]{cumulant generating function}
\acrodef{CVS}[CVS]{cardiovascular system}
\acrodef{CIR}[CIR]{channel impulse response}
\acrodef{FPT}[FPT]{first passage time}
\acrodef{IoBNT}[IoBNT]{Internet of Bio-Nano Things}
\acrodef{IG}[IG]{inverse Gaussian}
\acrodef{ISI}[ISI]{inter-symbol-interference}
\acrodef{MGF}[MGF]{moment generating function}
\acrodef{MIGHT}[MIGHT]{mixture of inverse Gaussians for hemodynamic transport}
\acrodef{MIMO}[MIMO]{multiple-input multiple-output}
\acrodef{MISO}[MISO]{multiple-input single-output}
\acrodef{SIMO}[SIMO]{single-input multiple-output}
\acrodef{SISO}[SISO]{single-input single-output}
\acrodef{PDF}[PDF]{probability density function}
\acrodef{RV}[RV]{random variable}
\acrodef{SNR}[SNR]{signal-to-noise ratio}
\acrodef{Tx}[Tx]{transmitter}
\acrodef{Rx}[Rx]{receiver}
\acrodef{UCA}[UCA]{uniform concentration assumption}
\acrodef{VN}[VN]{vessel network}
\acrodef{wrt}[w.r.t.]{with respect to}
\acrodefplural{IG}[IGs]{inverse Gaussians}
\acrodefplural{PDF}[PDFs]{probability density functions}
\acrodefplural{RV}[RVs]{random variables}
\acrodefplural{SNR}[SNRs]{signal-to-noise ratios}
\acrodefplural{GFP}[GFPs]{green fluorescent proteins}
\acrodefplural{CIR}[CIRs]{channel impulse responses}
\acrodefplural{VN}[VNs]{vessel networks}

\begin{document}

\title{%
Mixture of Inverse Gaussians for Hemodynamic Transport (MIGHT) in Multiple-Input Multiple-Output Vascular Networks
\vspace*{-2mm}
\thanks{$^{*}$Co-last authorship.}
\thanks{This work was funded in part by the German Federal Ministry of Research, Technology and Space (BMFTR) through Project Internet of Bio-Nano-Things (IoBNT), in part by the German Research Foundation (Deutsche Forschungsgemeinschaft, DFG) under GRK 2950 -- ProjectID 509922606 and under grant number SCHA 2350/2-1, in part by the European Union’s Horizon Europe -- HORIZON-EIC-2024-PATHFINDEROPEN-01 under grant agreement Project N. 101185661, and in part by the Horizon Europe Marie Skodowska Curie Actions (MSCA)-UNITE under Project 101129618.}
}

\author{\IEEEauthorblockN{\scalebox{.94}{%
  Timo Jakumeit$^1$, Bastian Heinlein$^{1,2}$, Nunzio Tuccitto$^3$, Robert Schober$^1$, Sebastian Lotter$^{1,*}$, and Maximilian Sch\"afer$^{1,*}$%
}}\\[-0.4cm]
\thanks{Parts of this paper were previously published in a conference version~\cite{Jakumeit2025a}.}
\IEEEauthorblockA{\small $^1$ Friedrich-Alexander-Universität Erlangen-Nürnberg (FAU), Erlangen, Germany\\
$^2$ Technical University of Darmstadt, Darmstadt, Germany\\
$^3$ University of Catania, Catania, Italy}\vspace*{-9mm}
}

\maketitle

\begin{abstract}
\Ac{MC} in the \ac{CVS} is a key enabler for many envisioned medical applications inside the human body, such as targeted drug delivery, early disease detection, and continuous health monitoring.
The design of synthetic \ac{MC} systems for such applications requires suitable models for the signaling molecule propagation through complex \acp{VN}.
Existing theoretical models offer limited analytical tractability and lack closed-form solutions, making the analysis of realistic large-scale \acp{VN} either infeasible or not insightful.
To overcome these limitations, in this paper, we propose a novel closed-form physical model, termed \ac{MIGHT}, for the advection-diffusion-driven transport of signaling molecules through complex \acp{VN}.
The model represents the received molecule flux as a weighted sum of \ac{IG} distributions, parameterized by the physical properties of the underlying \ac{VN}. We show that \ac{MIGHT} is capable of accurately representing the transport dynamics of signaling molecules in complex large-scale \acp{VN} ranging from simple \ac{SISO} to complex \ac{MIMO} network topologies. The accuracy of the proposed model is validated by comparison to the results from an existing convolution-based \ac{VN} model and numerical finite-element simulations, with all finite-element simulation data openly available on Zenodo.
Furthermore, we investigate three applications of the proposed model, namely (i) the reduction of large \ac{SISO}-\acp{VN} to obtain simplified representations preserving the essential transport dynamics, (ii) the identification and analysis of network regions that are most important for molecule transport in \ac{MIMO}-\acp{VN} comprising multiple \acp{Tx} and multiple \acp{Rx}, and (iii) the estimation of representative \ac{SISO}-\acp{VN} that can reproduce the received signal of an unknown \ac{SISO}-\ac{VN}.
\end{abstract}

\acresetall


\section{Introduction}\label{sec:Introduction}
Located at the intersection of communications engineering and life sciences, \ac{MC} investigates the exchange of information via signaling molecules between diverse entities, including biological systems, synthetic sensors, and engineered devices.
\ac{MC} is expected to provide critical insights and tools for the realization of various innovative medical applications. These include the broader vision of the \ac{IoBNT}, which underpins specific use cases such as early cancer detection and localization, as well as targeted drug delivery~\cite{Akyildiz2015, Mosayebi2019, ChudeOkonkwo2017}.
Most of the medical applications of \ac{MC} are expected to operate inside or to interface with the human body~\cite{Felicetti16}. Therefore, the \ac{CVS} is one of the main envisioned application domains for synthetic \ac{MC} systems, due to its pervasiveness in the human body and its significance for vital physiological processes. However, in order to realize synthetic \ac{MC} systems that sucessfully operate inside the \ac{CVS}, it is crucial to derive insightful channel models capable of characterizing the propagation of signaling molecules in the complex \acp{VN} constituting the \ac{CVS} \cite{Jamali2019}. 

While early works in \ac{MC} mainly focused on investigating signaling molecule propagation within isolated vessels, see, e.g., \cite{Schafer2021,Felicetti14,Zoofaghari19,Schäfer19,Wicke18,Lo19}, it has become clear that a comprehensive understanding of complex \acp{VN} is essential. 
This is particularly true because the macroscopic topology of a \ac{VN}, which governs flow distribution and path availability, typically has a far greater impact on signaling molecule transport than microscopic vessel-specific effects such as detailed flow profiles or molecule-wall interactions. 
To this end, physically accurate and insightful \textit{network-level} models for \acp{VN} are indispensable for both the investigation of the signaling molecule transport inside the \ac{CVS} and the design of future synthetic \ac{MC} systems for in-body environments. 

In particular, the development of insightful, analytically tractable, and closed-form models is desirable for several reasons. 
In communication-oriented applications of \ac{MC}, such channel models are essential for the derivation of capacity, mutual information, modulation schemes, and optimal detectors~\cite{Rose19,Pierobon16,Kuran21}. 
Likewise, for medical applications of \ac{MC}, including targeted drug delivery and continuous health monitoring, analytical channel models that provide physical insight are of critical importance \cite{Felicetti16,Ghavami20,ChudeOkonkwo2017}. 
For instance, in early cancer detection and localization, source inference relies on model inversion, which in turn requires tractable analytical models to ensure solvability and uniqueness of the solution.
Moreover, since \acp{VN} in in-body environments can be large-scale and topologically complex, computationally efficient and analytically tractable models are indispensable for both the analysis of signaling molecule propagation and the design of efficient communication, localization, and estimation schemes within such networks.

Previously, several works addressed channel modeling for \ac{MC} in \acp{VN}. In~\cite{Chahibi2013}, an analytical model for advective-diffusive molecule transport in \acp{VN} under time-varying flow was derived and subsequently extended in~\cite{Chahibi2014, Chahibi2015} by channel capacity analysis and additional transport phenomena. 
While the model captures relevant environmental effects, it is analytically unwieldy due to the involvement of infinite-dimensional matrices, inversions, and convolutions. Moreover, the analysis in \cite{Chahibi2013,Chahibi2014, Chahibi2015} is limited to tree-like \acp{VN}, neglecting multi-path dynamics from signal splitting and recombination at vessel bifurcations and junctions.
In~\cite{Mosayebi2019}, a \ac{1-D} \ac{VN} model is employed for early cancer detection in the \ac{CVS}. While the study considers multi-path dynamics, it relies on the assumption of a quasi-steady-state signaling molecule concentration inside the \ac{VN} and is not suitable for time-varying molecule releases.
In~\cite{Gomez2021}, a steady-state, discrete-time, advection-only Markov chain model was proposed to describe nanobot transport in a simplified circulatory \ac{VN} of the human \ac{CVS}, capturing branching probabilities and flow directionality, but omitting temporal signal resolution.
In~\cite{Tjabben2024}, a graph-based model was proposed for  predicting the propagation of degrading microbubbles in a closed-loop \ac{VN}. While analytically simple, the model is not validated through simulations or experiments, limiting its predictive utility.
In~\cite{Jakumeit2025}, we introduced a \ac{1-D} model for advective-diffusive molecule transport in \acp{VN}, validated through COMSOL Multiphysics\textsuperscript{\textregistered} simulations and simple experiments. However, the model involves convolutions that lack closed-form solutions and require numerical evaluation. As the number of convolutions in the model from \cite{Jakumeit2025} grows linearly with both the number of paths and the number of vessels per path, analyzing large-scale \acp{VN}, as commonly found in the \ac{CVS}, is computationally demanding and yields limited analytical insights. 
We extended this model to additionally include reversible sorption at the channel walls in~\cite{Jakumeit2025b} and validated it thoroughly through testbed experiments. The extended model provides accurate predictions, however, it still suffers from analytical unwieldiness due to convolutions and integrals that lack closed-form solutions.
Moreover, none of the existing channel models for \acp{VN} \cite{Chahibi2013,Chahibi2014, Chahibi2015,Mosayebi2019,Tjabben2024,Jakumeit2025,Jakumeit2025b} explicitly accounts for \ac{MIMO} characteristics, i.e., \acp{VN} with multiple flow inlets and outlets (Flow-\ac{MIMO}), which are typical when focusing on subsystems within the \ac{CVS}, or systems involving multiple simultaneously operating \acp{Tx} and \acp{Rx} (Com-\ac{MIMO}) embedded in the same \ac{VN}.
In general, both aspects may occur jointly, giving rise to \acp{VN} that feature multiple flow inlets and outlets as well as multiple \acp{Tx} and multiple \acp{Rx} within the same vasculature (here referred to as \ac{MIMO}-\ac{VN}).

In summary, existing \ac{VN} channel models tend to trade off analytical tractability, physical accuracy, and scalability, without achieving all three, and mostly do not capture the full topological complexity of relevant \acp{VN}. In this paper, we aim to overcome the limitations of existing channel models for \acp{VN} by proposing a novel closed-form channel model for advective-diffusive \ac{MC} in \acp{VN}, termed \textit{\ac{MIGHT}}. The proposed model overcomes the limitations of exiting models by offering a physically grounded, analytically tractable framework suitable for the analysis and design of \ac{MC} systems operating in topologically complex, large-scale \ac{MIMO}-\acp{VN}. The proposed \ac{MIGHT} model extends the well-known observation that the \ac{FPT} in advective-diffusive channels follows an \ac{IG} distribution~\cite{Srinivas2012,Haselmayr2017,Lin2016,Li2014} from individual vessels to \acp{VN}. Ultimately, the received signal in arbitrarily complex \acp{VN} comprising multiple transport paths can be expressed as a finite sum of weighted \acp{IG}, parameterized by the physical properties of the \ac{VN}.

The main contributions of this work can be summarized as follows:
\begin{enumerate}
    \item We derive a novel, closed-form, and analytically tractable model for advective-diffusive hemodynamic molecule transport in \ac{MIMO}-\acp{VN}, based on mixtures of \acp{IG}. The model is validated for both \ac{SISO}- and \ac{MIMO}-\ac{VN} topologies through a comparison with the model in~\cite{Jakumeit2025} and numerical \ac{3-D} finite-element simulations in COMSOL. 
    \item For the first time in the \ac{MC} literature, we analyze the signaling molecule transport in large-scale and topologically complex \acp{VN} containing more than 400 distinct transport paths through the \ac{VN} and multiple \acp{Tx} and \acp{Rx}, highlighting the model’s ability to handle \ac{MIMO}-\acp{VN} of large size and high complexity.
    \item Based on the proposed \ac{MIGHT} model, we derive a method for the structural reduction of \ac{SISO}-\acp{VN} to simplified representations that preserve the main transport dynamics.
    \item We further derive a vessel importance metric that provides an intuitive means for quantifying the relative importance of different regions in large-scale \ac{MIMO}-\acp{VN} with regards to signaling molecule transport.
    \item For \ac{SISO}-\acp{VN}, we propose an algorithm for the estimation of representative \acp{VN} that can reproduce the received signal of an unknown \ac{VN}.
\end{enumerate}

Compared to the conference version~\cite{Jakumeit2025a} of this paper, the proposed model goes beyond the modeling of \acp{VN} with a single flow inlet and outlet and a single \ac{Tx}-\ac{Rx} pair, and captures the signaling molecule dynamics in \acp{VN} of arbitrary size and complexity with multiple flow inlets and outlets (Flow-\ac{MIMO}-\acp{VN}). Moreover, while the model in~\cite{Jakumeit2025a} only allowed for the \ac{Tx} to be located at the network inlet and the \ac{Rx} to be in the network outlet pipe, the proposed model allows for an arbitrary number of \acp{Tx} and \acp{Rx} at \textit{arbitrary locations} in the \ac{MIMO}-\ac{VN}. Additionally, compared to \cite{Jakumeit2025a}, we describe the proposed methods for the reduction of \acp{VN} and the representative \ac{VN} estimation based on the received signal in more detail, and propose additional metrics for quantifying the importance of different \ac{VN} regions.

The remainder of this paper is structured as follows: Section~\ref{sec:SystemModel} introduces the system model, including a comprehensive description of \ac{VN} topologies, forming the basis for the derivation and validation of the proposed \ac{MIGHT} model in Section~\ref{sec:IGD}. 
Moreover, Section~\ref{sec:IGD} validates the proposed model by the analysis of several \ac{SISO}- and \ac{MIMO}-\acp{VN} and a comparison to results obtained from the numerical model in~\cite{Jakumeit2025} and finite-element simulations, respectively. 
Section~\ref{sec:NumericalResults} introduces three applications of the proposed model, i.e., the structural reduction of \ac{SISO}-\acp{VN}, vessel importance scoring in \ac{MIMO}-\acp{VN}, and estimation of representative \ac{SISO}-\acp{VN} from the received signal of an unknown \ac{VN}. Finally, Section~\ref{sec:Conclusion} concludes the paper and addresses several topics for future work. Extended proofs of the derivations in Section~\ref{sec:IGD} can be found in the appendix.

\section{System Model}\label{sec:SystemModel}
Below, we introduce the system model.
First, we extend the formal definition of \ac{SISO}-\acp{VN} in~\cite{Jakumeit2025} to accommodate the more general case of \ac{MIMO}-\acp{VN}.
Second, we introduce models for \acp{Tx} and \acp{Rx} operating in \ac{MIMO}-\acp{VN}.
Third, we detail the mechanisms governing molecule transport in the \ac{MIGHT} model, namely advection and diffusion.

\subsection{Vessel Network Definition}\label{ssec:VN_Definition}
\begin{figure*}
    \centering
    \includegraphics[width=\linewidth]{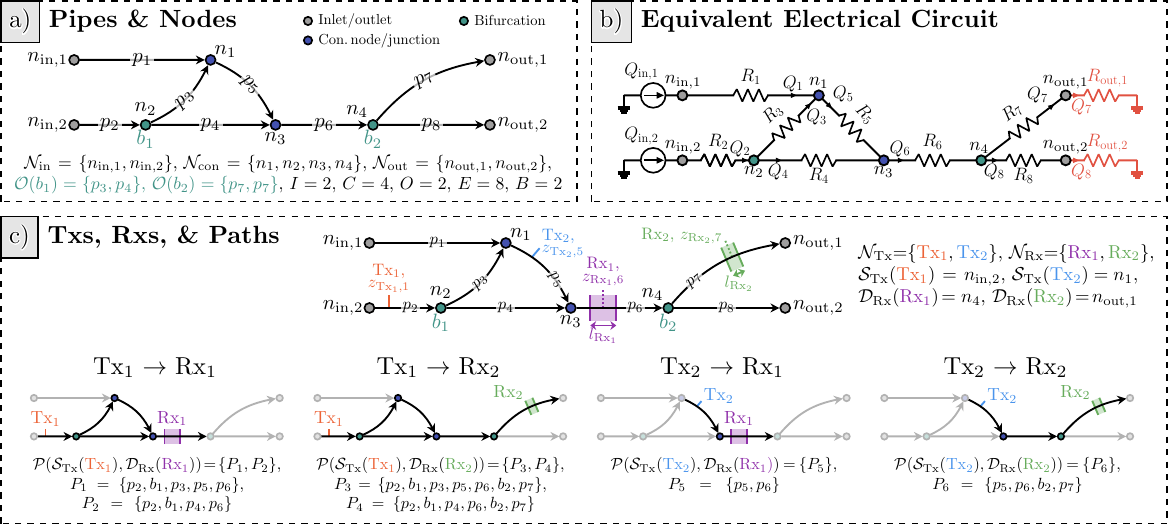}
    \caption{\textbf{Notation for \ac{VN} topology, equivalent electrical circuits, \acp{Tx}/\acp{Rx}, and paths.} \textbf{a)} Exemplary \ac{MIMO}-\ac{VN} with corresponding notation for the nodes (including inlets, connecting nodes, and bifurcations) and pipes. \textbf{b)} Equivalent electrical circuit for the \ac{VN}. At inlets $n_{\mathrm{in},a}\in \mathcal{N}_\mathrm{in}$, different inflow rates $Q_{\mathrm{in},a}$ can be applied. Terminal resistances $R_{\mathrm{out},b}$ are marked in red. Each pipe $p_i$ has an associated hydraulic resistance $R_i$. \textbf{c)} Multiple \acp{Tx} and \acp{Rx} with varying lengths placed at arbitrary positions in the exemplary \ac{VN}. The sets of paths between any \ac{Tx}-\ac{Rx} pair and related notation are given below the \ac{VN}.}
    \label{fig:Notation_and_Equiv_Circuit}
\end{figure*}

To model molecule transport in \acp{VN}, as found in the \ac{CVS}, the \ac{VN} topology is commonly approximated~\cite{Chahibi2013,Jakumeit2025} using three segment types (see Fig.~\ref{fig:Notation_and_Equiv_Circuit}a)):
\begin{enumerate}
    \itemsep0em
    \item \textit{Pipe:} A pipe $p_i$ is a cylindrical vessel transporting fluid from its inlet to its outlet, defined by its length $l_i$ and radius $r_i$. Pipes may connect to other pipes, bifurcations, or junctions at both ends. We denote the number of pipes contained in a \ac{VN} by $E$.
    \item \textit{Bifurcation:} A bifurcation $b_m$ is a \ac{0-D} connection, where one or more inflow pipe(s) split(s) into multiple outflow pipes. We denote the set of its outflow pipes by $\mathcal{O}(b_m)$. Bifurcations must be connected to pipes on both sides. We denote the number of bifurcations contained in a \ac{VN} by $B$. A bifurcation with multiple inflow and multiple outflow pipes is also referred to as a \textit{crossing}, see Fig.~\ref{fig:System_Model}a).
    \item \textit{Junction:} A junction is a \ac{0-D} connection, where multiple inflow pipes merge into one outflow pipe. Junctions must be connected to pipes on both sides, see Fig.~\ref{fig:Notation_and_Equiv_Circuit}a) and Fig.~\ref{fig:System_Model}a).
\end{enumerate}
Bifurcations, junctions, \ac{VN} inlet(s) and outlet(s), and any point that connects two pipes, are modeled as nodes. We distinguish between three types of nodes, see Fig.~\ref{fig:Notation_and_Equiv_Circuit}a) and Fig.~\ref{fig:System_Model}a):
\begin{enumerate}
\itemsep0em
    \item \textit{Inlet node:} Inlet nodes exist at the points of the \ac{VN} where fluid flow is introduced into the \ac{VN}. \acp{VN} may contain $I \in \mathbb{N}$ inlets, where $\mathbb{N}$ denotes the set of natural numbers. The set of inlet nodes is denoted by $\mathcal{N}_\mathrm{in}=\{n_{\mathrm{in},1},\ldots ,n_{\mathrm{in},I}\}$.
    \item \textit{Outlet node:} Outlet nodes exist at the points of the \ac{VN} where fluid flow leaves the \ac{VN}. \acp{VN} may contain $O \in \mathbb{N}$ outlets. The set of outlet nodes is denoted by $ \mathcal{N}_\mathrm{out}=\{n_{\mathrm{out},1},\ldots ,n_{\mathrm{out},O}\}$.
    \item \textit{Connecting node:} All other $C\in\mathbb{N}$ points in the \ac{VN} where pipes are connected to one another are referred to as connecting nodes. The set of connecting nodes is denoted by $\mathcal{N}_\mathrm{con}=\{n_{1},\ldots ,n_{C}\}$. Bifurcations and junctions belong to this class of nodes.
\end{enumerate}
Pipes are represented as directed edges between nodes, aligned with the direction of fluid flow, as determined in Subsection~\ref{ssec:Equivalent_Circuit_Theory}.
For any node type, the nodes at the inlet and outlet of a pipe $p_i$, i.e., its \textit{source node} and \textit{destination node}, are denoted by $\mathcal{S}(p_i)$ and $\mathcal{D}(p_i)$, respectively.

The representation based on nodes and directed edges allows any \ac{VN} to be described as a directed multigraph, cf.~Fig.~\ref{fig:System_Model}a).
The set of all distinct directed paths between a given (inlet/connecting) node $n_{a}\in\mathcal{N}_\mathrm{in}\cup \mathcal{N}_\mathrm{con}$ and another (connecting/outlet) node $n_{b}\in\mathcal{N}_\mathrm{con}\cup \mathcal{N}_\mathrm{out}$ is denoted by $\mathcal{P}(n_{a}, n_{b})$.
Each path $P_k$ comprises a subset of pipes and bifurcations given by
\begin{equation}\label{eqn:PathSet}
P_k = \left\lbrace p_i \mid i \in \mathcal{E}_k \right\rbrace \cup \left\lbrace b_m \mid m \in \mathcal{B}_k \right\rbrace,
\end{equation}
where $\mathcal{E}_k \subseteq \left\lbrace 1,\ldots ,E \right\rbrace$ and $\mathcal{B}_k \subseteq \left\lbrace 1,\ldots ,B \right\rbrace$ are the index sets of the pipes and bifurcations\footnote{Note that potential bifurcations at $n_a$ or $n_b$ are \textit{not} included in the path set in~\eqref{eqn:PathSet}. This is because molecules in any path between $n_a$ and $n_b$ do not actually travel \textit{through} $n_a$ or $n_b$, but rather start at $n_a$ and end at $n_b$.} included in $P_k$. Any path must contain at least two pipes, i.e., $|\mathcal{E}_k|>1$, where $|\cdot|$ denotes the cardinality of a set. In path $P_k$, we explicitly denote the pipes at the beginning and the end of the path as $p_q$ and $p_w$, respectively, with $q,w\in\mathcal{E}_k$.

Throughout this paper, we distinguish between Flow-\ac{MIMO} (multiple flow inlets/outlets) and Com-\ac{MIMO} (multiple \acp{Tx} and \acp{Rx}) systems. Unless stated otherwise, “\ac{MIMO}-\ac{VN}” refers to the general case combining both effects.
The topology-related notation is illustrated for an exemplary \ac{MIMO}-\ac{VN} in Figs.~\ref{fig:Notation_and_Equiv_Circuit}a) and ~\ref{fig:Notation_and_Equiv_Circuit}c). An overview of the introduced notation is given in Table~\ref{tab:Notation}.

\subsection{Transmitters and Receivers}\label{ssec:Txs_and_Rxs}
In the proposed \ac{1-D} model, \acp{Tx} are \ac{0-D} points where signaling molecules are released. In particular, the set of $U\in\mathbb{N}$ \acp{Tx} in a \ac{VN} is denoted by $\mathcal{N}_\mathrm{Tx}=\{\mathrm{Tx}_1,\ldots,\mathrm{Tx}_U \}$. 
The source node of the pipe containing $\mathrm{Tx}_g$ is denoted as $\mathcal{S}_\mathrm{Tx}(\mathrm{Tx}_g)$.
The longitudinal position of $\mathrm{Tx}_g$ in pipe $p_i$ is denoted by $z_{\mathrm{Tx}_g}\in[0,l_i]$. The number of molecules released at $\mathrm{Tx}_g$ is denoted by $N_g$. 
For each \ac{Tx}, different injection functions over $t$ are possible. The injection function of $\mathrm{Tx}_g$ is given as $f_{\mathrm{Tx}_g}(t)$ with $\int_0^\infty f_{\mathrm{Tx}_g}(t)\,\mathrm{d}t=N_g$, see Fig.~\ref{fig:System_Model}b).

The set of $V\in\mathbb{N}$ transparent\footnote{We restrict our analysis to transparent \acp{Rx} in this work to focus on the channel behavior. Incorporating more realistic \ac{Rx} models, including absorbing \acp{Rx}, is left for future work.} \acp{Rx} in a \ac{VN} is denoted by $\mathcal{N}_\mathrm{Rx}=\{\mathrm{Rx}_1,\ldots,\mathrm{Rx}_V \}$. Any $\mathrm{Rx}_h$, placed in pipe $p_i$, is characterized by its length $l_{\mathrm{Rx}_h}\in(0,l_i]$ and its longitudinal center position $z_{\mathrm{Rx}_h}\in[0+l_{\mathrm{Rx}_h}/2,l_i-l_{\mathrm{Rx}_h}/2]$, i.e., the \ac{Rx} domain spans $z\in [z_{\mathrm{Rx}_h}-l_{\mathrm{Rx}_h}/2,z_{\mathrm{Rx}_h}+l_{\mathrm{Rx}_h}/2]$ in $p_i$, see Figs.~\ref{fig:Notation_and_Equiv_Circuit}c) and \ref{fig:System_Model}b). The destination node of the pipe, in which $\mathrm{Rx}_h$ is located, is denoted by $\mathcal{D}_\mathrm{Rx}(\mathrm{Rx}_h)$.

All \ac{Tx}- and \ac{Rx}-related notation is summarized in Table~\ref{tab:Notation} and illustrated for an exemplary \ac{MIMO}-\ac{VN} in Fig.~\ref{fig:Notation_and_Equiv_Circuit}c).

\subsection{Advective Molecule Transport}\label{ssec:Equivalent_Circuit_Theory}
\begin{figure*}
    \centering
    \includegraphics[width=\linewidth]{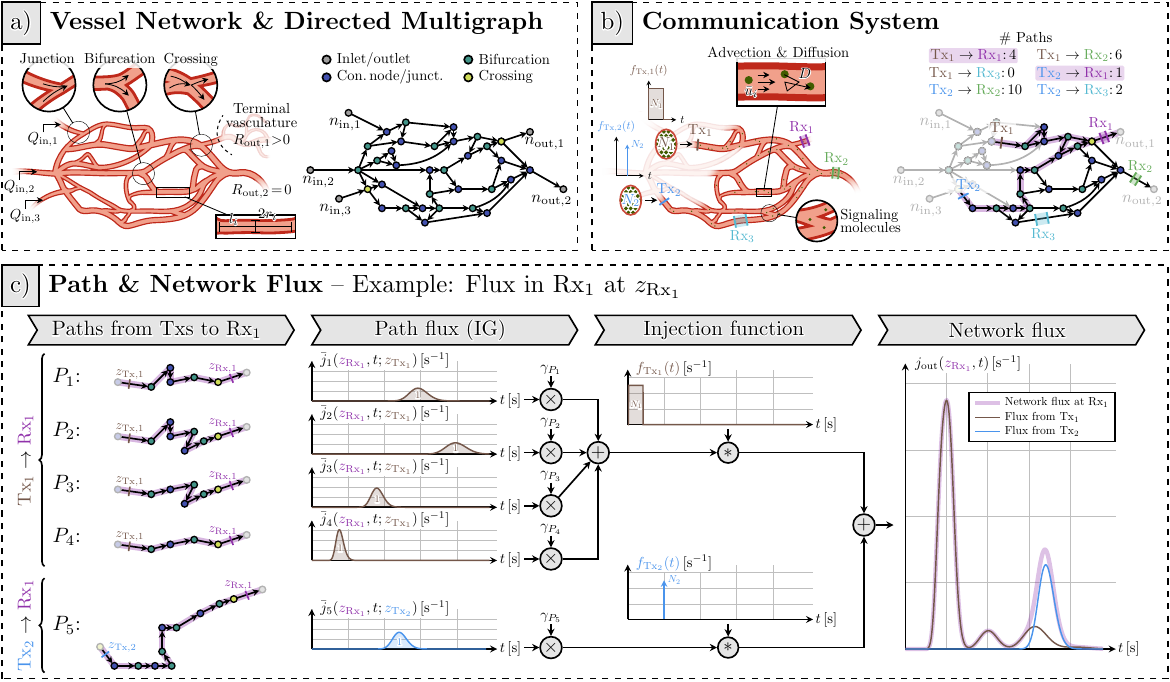}
    \caption{\textbf{System and channel model.} \textbf{a) Vessel network \& directed multigraph:} An exemplary \ac{VN} with multiple flow inlets (with inflow rates $Q_{\mathrm{in},a}$) and outlets (with terminal resistances representing the terminal vasculature) is shown. Vessel segments, i.e., pipes, are defined by their lengths $l_i$ and radii $r_i$ and are connected via bifurcations (crossings) and junctions. Following the fluid flow direction, the \ac{VN} is represented as a directed multigraph in \ac{1-D}. 
    \textbf{b) Communication system:} Multiple \acp{Tx} and \acp{Rx}, placed at arbitrary positions in the \ac{VN}, form a \ac{MIMO} communication system. Each $\mathrm{Tx}_g$ releases $N_g$ signaling molecules according to injection function $f_{\mathrm{Tx}_g}(t)$ and uniformly in the pipe cross-section. Molecules propagate via advection and diffusion through multiple possible paths from the \acp{Tx} to the \acp{Rx}. Each $\mathrm{Rx}_h\in\mathcal{N}_\mathrm{Rx}$ may span a different spatial extent, characterized by its length $l_{\mathrm{Rx}_h}$ and center position $z_{\mathrm{Rx}_h}$. For communication purposes, all parts of the \ac{VN} upstream of the \acp{Tx} can be neglected, though they remain relevant for flow calculations. 
    \textbf{c) Path \& network flux:} As an example, the network flux in $\mathrm{Rx}_2$ at $z_{\mathrm{Rx}_2}$ is determined by considering all contributing \acp{Tx} and the respective paths to $\mathrm{Rx}_2$. Each path $P_k$ contributes a path flux, weighted by its path fraction $\gamma_{P_k}$, which is governed by flow rate fractions at the bifurcations in the respective path. For each \ac{Tx}, the weighted path fluxes are summed and convolved with its injection function. The final network flux is obtained by summing these contributions across all \acp{Tx}.}
    \label{fig:System_Model}
\end{figure*}

The proposed \ac{MIGHT} model captures advection- and diffusion-driven molecule transport, see Fig.~\ref{fig:System_Model}b). Below, we first model advective transport.

At each inlet node $n_{\mathrm{in},a}\in\mathcal{N}_\mathrm{in}$, an arbitrary flow rate $Q_{\mathrm{in},a}>0$ is applied.
This induces a time-invariant fluid flow\footnote{Time-invariant blood flow is an accurate approximation in medium-sized and small vessels of the \ac{CVS}, where pulsatility is damped by the Windkessel effect of the preceding large arteries~\cite{Aaronson2012}.} in each pipe $p_i$ of the \ac{VN}, characterized by the flow rate $Q_i$ and the cross-sectional average flow velocity
\begin{equation}\label{eqn:Average_Cross-Sectional_Velocity}
\bar{u}_i = \dfrac{Q_i}{\pi r_i^2}\,.
\end{equation}
The flow rates and velocities in~\eqref{eqn:Average_Cross-Sectional_Velocity} are computed using an equivalent electrical circuit model, similar to~\cite{Jakumeit2025} and~\cite{Chahibi2013}. Specifically, each pipe $p_i$ is assigned a hydraulic resistance according to the Hagen–Poiseuille law~\cite[Eq.~(14)]{Chahibi2013}
\begin{equation}\label{eqn:HydraulicResistance}
R_i = \dfrac{8\mu l_i}{\pi r_i^4}\,,
\end{equation}
where $\mu$ denotes the dynamic fluid viscosity. Fluid inflow sources are modeled as electrical current sources, and a circuit mirroring the \ac{VN} topology is constructed, see Figs.~\ref{fig:Notation_and_Equiv_Circuit}a) and~\ref{fig:Notation_and_Equiv_Circuit}b).
At the outlet nodes $n_{\mathrm{out},b}\in\mathcal{N}_\mathrm{out}$, non-zero \textit{terminal hydraulic resistances} may be present; hence, zero outlet pressure is not assumed by default. 
These terminal resistances are modeled as additional resistors $R_{\mathrm{out},b}$ connected to electrical ground in the equivalent circuit. 
Physiologically, they represent the hydraulic resistance of the downstream vasculature that is not explicitly included in the considered \ac{VN}, see Fig.~\ref{fig:System_Model}a). 
By applying node voltage analysis, the electrical currents, and thus the flow rates $Q_i$, are then obtained directly.
The equivalent circuit corresponding to the exemplary \ac{MIMO}-\ac{VN} in Fig.~\ref{fig:Notation_and_Equiv_Circuit}a) is depicted in Fig.~\ref{fig:Notation_and_Equiv_Circuit}b), with terminal resistances highlighted in red.

\subsection{Diffusive Molecule Transport}\label{ssec:DiffusiveMoleculeTransport}

In addition to advective transport, molecules propagate through the \ac{VN} via diffusion.
Under the assumption of the Aris–Taylor regime, the effective diffusion coefficient in pipe $p_i$ is given by~\cite[Eq.~(26)]{Aris1956}
\begin{equation}\label{eqn:ArisTaylorEffectiveDiffusionCoefficient}
\bar{D}_i = \dfrac{r_i^2 \bar{u}_i^2}{48 D} + D
\end{equation}
and captures the combined effects of molecular diffusion, characterized by the molecular diffusion coefficient $D$, and shear-induced dispersion resulting from the non-uniform velocity profile across the pipe cross-section. 
The validity of the Aris-Taylor regime assumption in~(\ref{eqn:Average_Cross-Sectional_Velocity}) and~(\ref{eqn:ArisTaylorEffectiveDiffusionCoefficient}), and the resulting \ac{1-D} modeling, was previously confirmed in~\cite{Jakumeit2025, Jakumeit2025b} and is further supported by the numerical results in Section~\ref{ssec:Model_Validation}.

\section{Inverse Gaussian Mixture Channel Model}\label{sec:IGD}
Based on the concept of the \ac{FPT}, in the following, we derive expressions for the molecule flux in both single pipes and paths comprising multiple pipes. An exact solution is obtained for a special case, while for general paths, we find an approximation whose error scales inversely with the Péclet number and which is therefore negligible in the considered regime. Lastly, we generalize the model to \ac{MIMO}-\acp{VN}, which includes as special cases \ac{SISO}-, \ac{SIMO}-, and \ac{MISO}-\acp{VN}.

\subsection{Single Pipe Model}\label{ssec:SinglePipeModel}

When modeling molecule transport by diffusion and flow through a pipe $p_i$, most works in \ac{MC} assume that molecules are already present at time $t = 0$, reflected by the initial condition $c(z_i,t=0) = N\delta(z_i)$ \cite{Jakumeit2025,Jamali2019}, where $c(z_i,t)$, $N$, and $\delta (\cdot)$ denote the molecule concentration at time $t$ and longitudinal position $z_i\in [0,l_i]$ within $p_i$, the number of injected molecules, and the Dirac delta function, respectively. In practice, however, molecules typically enter $p_i$ at $t = 0$, e.g., from the preceding pipe, by syringe injection, or by release from nanodevices. Accordingly, in this work, we adopt the more realistic assumption of molecule injection at the pipe inlet $z_i = 0$ at $t = 0$, see~\eqref{eq:bc} in the appendix.
We define the \ac{FPT} as a \ac{RV} describing the time at which a molecule, entering $p_i$ at the pipe inlet at $t = 0$, first reaches position $z_i$, i.e.,
\begin{align} T_i(z_i) = \mathrm{inf}\{t >0: z_i(T_i) = z_i(t)\,\vert\, z_i(t_0) = 0\}, \label{eq:fpt} \end{align}
where $z({t_0})$ and $z(T_i)$ denote the molecule’s positions at $t = 0$ and $t = T_i$, respectively.
\begin{theorem}\label{theorem1}
The \ac{FPT} $T_i(z_i)$ of a single molecule entering pipe $p_i$ at $t = 0$ is \ac{IG}-distributed, i.e.,
\begin{align}
    T_i(z_i) \sim \mathrm{IG}(\mu_i(z_i),\theta_i(z_i)),
    \label{eq:igd}
\end{align}
with the mean, variance, and scale parameter given by
\begin{align}
    &\mu_i(z_i) = \frac{z_i}{\bar{u}_i}, 
    &\sigma_i^2(z_i) = \frac{2\bar{D}_i z_i}{\bar{u}_i^3},
    &&\theta_i(z_i) = \frac{\sigma^2_i(z_i)}{\mu_i(z_i)},
    \label{eq:pipe-moms}
\end{align}
respectively.
The molecule flux $j_i(z_i,t)$ in $\si{\per\second}$ at position $z_i$ in pipe $p_i$ is the \ac{PDF} $f_{\mathrm{IG}}(t,z;\mu,\theta)$ of $T_i(z_i)$ in \eqref{eq:igd}, i.e.,
\begin{align}\label{eq:pipe-pdf} 
    j_i(z_i,t) = f_{\mathrm{IG}}(t,z_i;\mu_i,\theta_i) &= \frac{\mu_i(z_i)}{\sqrt{2\pi\theta_i(z_i) t^3}}\mathrm{e}^{\left( -\frac{(t-\mu_i(z_i))^2}{2\theta_i (z_i) t}\right)}\,.
\end{align}

Proof: The proof is provided in Appendix~\ref{sec:proof1}. 
\end{theorem}

\subsection{Multiple Pipe Path Model}
\label{ssec:pathmodel}

Building on the model for the flux in a single pipe in Theorem~\ref{theorem1}, we derive the molecule flux for a path $P_k\in\mathcal{P}(n_a,n_b)$ consisting of multiple pipes, see Fig.~\ref{fig:System_Model}b). In particular, consider $P_k$ between nodes $n_a$ and $n_b$, with first pipe $p_q$, and last pipe $p_w$. We assume $\mathrm{Tx}_g$ is positioned at $z_q$ in pipe $p_q$. The flux is observed at position $z_w$ in pipe $p_w$.
Since the propagation in each pipe is modeled as an independent advection-diffusion process, the pipe \acp{FPT} $T_i(z_i)$ in path $P_k$ are mutually independent, i.e., $T_i(z_i)$ is unaffected by $T_{i'}(z_{i'})$, with $ i,i'\in\mathcal{E}_k,i\neq i'$. Hence, the path \ac{FPT} $\bar{T}_k(z_w;z_q)$ of a molecule injected at position $z_q$ in pipe $p_q$ at time $t = 0$, propagating through $P_k$ and reaching position $z_w$ in pipe $p_w$, is the sum of the individual pipe \acp{FPT}
\begin{align}
    \bar{T}_k(z_w;z_q)=\underbrace{T_q(l_q-z_q)}_{\text{First pipe}}+\underbrace{T_w(z_w)}_{\text{Last pipe}}+\hspace{-1mm}\underbrace{\sum_{i\in\mathcal{E}_k\backslash \{q,w\}}\hspace{-3mm} T_i(l_i)}_{\text{Intermediate pipes}},\label{eq:sumt} 
\end{align}
i.e., the path \ac{FPT} is obtained as the sum of the \ac{FPT} $T_q$ in the first pipe, the \acp{FPT} $T_i$ in all intermediate pipes, and the \ac{FPT} $T_w$ in the last pipe of the path.
Therefore, the path flux $\bar{j}_k(z_w,t;z_q)$ at $z_w$ in pipe $p_w$ can be obtained by convolving the pipe \acp{PDF} in~\eqref{eq:pipe-pdf}
\begin{align}
    \bar{j}_k(z_w,t;z_q)\hspace{-.5mm}=\hspace{-.5mm}
     j_q(l_q\hspace{-.5mm}-\hspace{-.5mm}z_q,t)\hspace{-.5mm}\ast\hspace{-.5mm} j_w(z_w,t)\hspace{-.5mm}\ast\hspace{-.5mm}\left( \Asterisk_{i\in\mathcal{E}_k\backslash\{q,w\}} j_i(l_i,t)\right).
     \label{eq:path-conv}
\end{align}
Here, $\Asterisk_{i\in\mathcal{E}_k\backslash\{q,w\}}$ denotes the temporal convolution of all pipe fluxes $j_i$ at their respective pipe outlets $l_i$ in path $P_k$, except for the first and last pipes $p_q$ and $p_w$, where the positional argument is adjusted according to the injection and readout points. Moreover, $\ast$ denotes a single convolution \ac{wrt} time. Below, we derive two closed-form expressions for \eqref{eq:path-conv}.

\begin{theorem}\label{theorem2}
    For the special case of homogeneous parameters in pipes $p_i\in P_k$, i.e., $\bar{u}_i = \bar{u}$ and $\bar{D}_i = \bar{D}$, the path \ac{FPT} $\tilde{T}_k(z_w;z_q)$ of a molecule injected by $\mathrm{Tx}_g$ at $z_q$ into the first pipe $p_q$ of the path at $t = 0$, and reaching position $z_w$ in the last pipe $p_w$ of the path, is again \ac{IG}-distributed
    \begin{align}
        \bar{T}_{k}(z_w;z_q)=\tilde{T}_k(z_w;z_q)\sim \mathrm{IG}(\tilde{\mu}_k(z_w;z_q), \tilde{\theta}_k(z_w;z_q)),
        \label{eq:tk-hom}
    \end{align}
    with the path mean, variance, and scale parameter given by
    \begin{align}
        \tilde{\mu}_k(z_w;z_q) &= \frac{\bar{l}_k(z_q,z_w)}{\bar{u}},\quad\nonumber
        \tilde{\sigma}_k^2(z_w;z_q) = \frac{2\bar{D}\bar{l}_k(z_q,z_w)}{\bar{u}^3},\nonumber\\
        \tilde{\theta}_k(z_w;z_q) &= \frac{\tilde{\sigma}_k^2(z_q,z_w)}{\tilde{\mu}_k(z_q,z_w)},
        \label{eq:pathMom}
    \end{align}
    where $\bar{l}_k(z_q,z_w)$ is the path length, defined by the injection position $z_q$ in the first pipe $p_q$, the lengths $l_i$ of the pipes lying between the first and last pipes of the path, and the readout position $z_w$ in the last pipe $p_w$, i.e.,
    \begin{align}\label{eqn:path_length}
        \bar{l}_k(z_q,z_w) =
        \underbrace{(l_q-z_q)}_{\text{First pipe}}+\underbrace{z_w}_{\text{Last pipe}}+\underbrace{\sum_{i\in\mathcal{E}_k\backslash\{q,w\}} l_i}_{\text{Intermediate pipes}}\,.
    \end{align}
    The \ac{PDF} of $\tilde{T}_k(z_w;z_q)$ is given as $\tilde{j}_k(z_w,t;z_q) = f_\mathrm{IG}(t,z_w;\tilde{\mu}_k(z_w;z_q), \tilde{\theta}_k(z_w;z_q))$, see~\eqref{eq:pipe-pdf}.
    
    Proof: The proof is provided in Appendix~\ref{sec:proof2}. 
\end{theorem}

For the general case of heterogeneous pipe parameters, where $l_i$, $r_i$, $\bar{u}_i$, and $\bar{D}_i$ vary across the pipes $p_i\in P_k$, the path \ac{FPT} $\bar{T}_k(z_w;z_q)$ at position $z_w$ in pipe $p_w$ is not exactly \ac{IG}-distributed. However, by applying the method of moment matching, $\bar{T}_k(z_w;z_q)$ can be approximated by an \ac{IG}-distributed \ac{RV} $\hat{T}_k(z_w;z_q)$, i.e.,
\begin{align}
    \bar{T}_k(z_w;z_q) \approx \hat{T}_k(z_w;z_q)\sim \mathrm{IG}(\hat{\mu}_k(z_w;z_q)\,,\hat{\theta}_k(z_w;z_q)),
    \label{eq:tk-het}
\end{align}
whose mean $\hat{\mu}_k(z_w;z_q)$ and variance $\hat{\sigma}_k^2(z_w;z_q)$ exactly match those of the true path \ac{FPT} $\bar{T}_k(z_w;z_q)$. Since all pipe \acp{FPT} in~\eqref{eq:sumt} are independent \acp{RV}, the mean $\bar{\mu}_k(z_w;z_q)$ and variance $\bar{\sigma}_k^2(z_w;z_q)$ of the true path \ac{FPT} $\bar{T}_k(z_w;z_q)$ are obtained by summing the means and variances of the individual pipe \acp{FPT}, which are matched to the moments of $\hat{T}_k(z_w;z_q)$ as follows
    \begin{align}
        \hat{\mu}_k(z_w;z_q)\hspace{-.5mm}&=\hspace{-.5mm} \bar{\mu}_k(z_w;z_q)\hspace{-.5mm}=\hspace{-.5mm}
        \mu_q(l_q\hspace{-.5mm}-\hspace{-.5mm}z_q)\hspace{-.5mm}+\hspace{-.5mm}\mu_w(z_w)\hspace{-.5mm}+\hspace{-.5mm}\hspace{-4mm}\sum_{i\in \mathcal{E}_k\backslash \left\lbrace q,w\right\rbrace}\hspace{-4mm} \mu_i(l_i),\label{eq:mom-het-mu}\\
        \hat{\sigma}^2_k(z_w;z_q) &\hspace{-.5mm}= \hspace{-.5mm}\bar{\sigma}_k^2(z_w;z_q)
        \hspace{-.5mm}=\hspace{-.5mm} \sigma_q^2(l_q\hspace{-.5mm}-\hspace{-.5mm}z_q)\hspace{-.5mm}+\hspace{-.5mm}\sigma_w^2(z_w)\hspace{-.5mm}+\hspace{-.5mm}\hspace{-4mm}\sum_{i\in \mathcal{E}_k\backslash \left\lbrace q,w\right\rbrace}\hspace{-4mm}\sigma_i^2(l_i),\label{eq:mom-het-sigma}
    \end{align} 
with 
\begin{align}
    \label{eq:mom-het-theta}\hat{\theta}_k(z_w;z_q)=\hat{\sigma}_k^2(z_w;z_q)/\hat{\mu}_k(z_w;z_q)\,.
\end{align}
Consequently, for heterogeneous path parameters, the path flux $\bar{j}_k(z_w,t;z_q)$ can be approximated by $\hat{j}_k(z_w,t;z_q)$, obtained from the \ac{PDF} in~\eqref{eq:pipe-pdf} with the matched moments in~\eqref{eq:mom-het-mu} and~\eqref{eq:mom-het-sigma}, i.e., $\bar{j}_k(z_w,t;z_q)\approx\hat{j}_k(z_w,t;z_q) = f_{\mathrm{IG}}(t,z_w; \hat{\mu}_k(z_w;z_q), \hat{\theta}_k(z_w;z_q))$.

\begin{theorem}\label{theorem3} 
For a given path $P_k$ with fixed pipe heterogeneity, see~\eqref{eq:vartheta}, the approximation in \eqref{eq:tk-het} is exact \ac{wrt} the first two moments and asymptotically exact \ac{wrt} the third moment for large Péclet numbers. 
In particular, the error $\Delta \xi = \bar{\xi}_k - \hat{\xi}_k$ in the skewness $\bar{\xi}_k$ of $\bar{T}_k$ and $\hat{\xi}_k$ of $\hat{T}_k$ follows as
    \begin{align}
        \Delta\xi \propto \tfrac{1}{\sqrt{\mathrm{Pe}_k}},
        \label{eq:app-err}
    \end{align}
where $\mathrm{Pe}_k$ is the path Péclet number, i.e., $\mathrm{Pe}_k = 2\hat{\mu}_k/\hat{\theta}_k$. 

Proof: The proof is provided in Appendix~\ref{sec:proof3}. 
\end{theorem}

The validity of the approximation $\bar{j}_k(z_w,t;z_q)\approx\hat{j}_k(z_w,t;z_q)$ is confirmed by comparison to the model in~\cite{Jakumeit2025} and finite-element simulations in COMSOL in Subsection~\ref{ssec:Model_Validation}. We note that moderately large $\mathrm{Pe}_k$ (e.g., $1\ll\mathrm{Pe}_k\ll 10^4$) are required for the Aris-Taylor regime to hold, in which case the molecule flux usually exhibits no skewness in $t$ and $\Delta \xi$ becomes negligibly small.

\subsection{MIMO Vessel Network Flux Model} 
\label{ssec:open_loop_vessel_network_flux_model}

Building upon the \ac{Tx} and \ac{Rx} models in Subsection~\ref{ssec:Txs_and_Rxs} and the path flux model derived in Subsection~\ref{ssec:pathmodel}, in the following, a general expression for the signaling molecule flux at any position in a \ac{MIMO}-\ac{VN} is presented. The flux is generally caused by several \acp{Tx} in the \ac{VN}.

In a first step, we derive the \ac{CIR} between a single $\mathrm{Tx}_g$ and position $z_i$ in pipe $p_i$. For $f_{\mathrm{Tx}_g}(t)=\delta (t)$, the \ac{CIR} in $\SI{}{\per\second}$ can be expressed as the flux observed at $z_i$ in pipe $p_i$ due to the release at $\mathrm{Tx}_g$, which in turn is obtained as the weighted sum of the path fluxes of all paths between $\mathrm{Tx}_g$ and pipe $p_i$, i.e.,
\begin{equation}\label{eqn:CIR_between_Tx_and_pi}
    h_{\mathrm{Tx}_g,i}(z_i,t;z_{\mathrm{Tx}_g})= \hspace*{-5mm}\sum_{\{P_k\in \mathcal{P}(\mathcal{S}_\mathrm{Tx}(\mathrm{Tx}_g),\mathcal{D}(p_i))\vert p_i\in P_k\}} \hspace*{-5mm}\gamma_{P_k} \bar{j}_k(z_i,t;z_{\mathrm{Tx}_g})\,.
\end{equation}
Here, the sum is over the set of all paths from the source node of $\mathrm{Tx}_g$ to the destination node of pipe $p_i$ that include $p_i$. Moreover, the fraction of molecules $\gamma_{P_k}$ propagating through path $P_k$ is obtained according to the law of mass conservation as~\cite[Eq.~(22)]{Mosayebi2019}   
\begin{equation}\label{eqn:PathWeight}
    \gamma_{P_k} = \prod\limits_{\substack{p_j,b_m\in P_k,\\ p_j\in\mathcal{O}(b_m )}} \dfrac{Q_{j}}{\sum_{p_v\in\mathcal{O}(b_m)} Q_{v}},
\end{equation}
with $\gamma_{P_k}\in[0,1]$, i.e., $\gamma_{P_k}$ is dictated by the fractions of flow rates at each bifurcation $b_m\in P_k$. Note that the \ac{CIR} in~\eqref{eqn:CIR_between_Tx_and_pi} is expressed in terms of the molecule flux, instead of molecule concentration as is often done in \ac{MC}. Hence, here, the \ac{CIR} is a \ac{PDF} over $t$, whereas a \ac{CIR} expressed in terms of normalized molecule concentration would be a \ac{PDF} over $z_i$.

Unlike the model in~\cite[Eq.~(8)]{Jakumeit2025}, \eqref{eqn:CIR_between_Tx_and_pi} requires no convolutions for the calculation of the \ac{CIR} and can instead be expressed as a sum of \acp{IG}. The sum provides a tractable and insightful closed-form expression for the molecule transport dynamics in \acp{VN}, in which the contributions of different paths originating from a given \ac{Tx} can be easily identified from the individual terms.
Equation~\eqref{eqn:CIR_between_Tx_and_pi} represents, to the best of our knowledge, the first closed-form expression for the \ac{CIR} in \acp{VN}.

In a second step, we use~\eqref{eqn:CIR_between_Tx_and_pi} to derive an expression for the total flux received at position $z_i$ in pipe $p_i$ due to arbitrary release functions at arbitrarily many \acp{Tx} in the \ac{VN} in $\SI{}{\per\second}$ as
\begin{equation}\label{eqn:network_flux_arbitrary_inj_functions}
    j_{\mathrm{out},i}(z_i,t) = \sum_{\mathrm{Tx}_g\in\mathcal{N}_\mathrm{Tx}}f_{\mathrm{Tx}_g}(t)\ast h_{\mathrm{Tx}_g,i}(z_i,t;z_{\mathrm{Tx}_g})\,.
\end{equation}
Note that the convolution in~\eqref{eqn:network_flux_arbitrary_inj_functions} vanishes for instantaneous injections at all \acp{Tx}, i.e., for $f_{\mathrm{Tx}_g}(t)=N_g\delta (t),\forall \mathrm{Tx}_g\in\mathcal{N}_\mathrm{Tx}$, in which case the total received flux is obtained as
\begin{equation}\label{eqn:network_flux_instant_inj_functions}
    j_{\mathrm{out},i}(z_i,t) = \sum_{\mathrm{Tx}_g\in\mathcal{N}_\mathrm{Tx}}N_g h_{\mathrm{Tx}_g,i}(z_i,t;z_{\mathrm{Tx}_g})\,.
\end{equation}

\subsection{Received Signal Model}

Based on the flux model in the preceding subsection, the received molecular signals can be derived.
By dividing~(\ref{eqn:network_flux_arbitrary_inj_functions}) by the flow velocity $\bar{u}_i$ in pipe $p_i$, the molecule concentration in $\si{\per\meter}$ at position $z_i$ in pipe $p_i$ is obtained as 
\begin{equation}\label{eqn:Network_Concentration}
    c_{\mathrm{out},i}(z_i,t)\approx \frac{1}{\bar{u}_i} j_{\mathrm{out},i}(z_i,t)\,.
\end{equation}
This assumes negligible diffusive flux at $z_i$ in $p_i$ compared to advective flux, which holds for $\mathrm{Pe}_k\gg 1$, i.e., in the Aris-Taylor regime.
This assumption is validated in Subsection~\ref{ssec:Model_Validation}.

Furthermore, the number of molecules observed in the domain of $\mathrm{Rx}_h$ positioned in pipe $p_i$ at center position $z_{\mathrm{Rx}_h}$ follows as
\begin{equation}\label{eqn:Nobs}
    N_{\mathrm{Rx}_h}(t)=\hspace{-1mm}\int_{z_{\mathrm{Rx}_h}-l_{\mathrm{Rx}_h}/2}^{z_{\mathrm{Rx}_h}+l_{\mathrm{Rx}_h}/2}\hspace{-15mm}c_{\mathrm{out},i}(z_i,t)\,\mathrm{d}z_i \approx l_{\mathrm{Rx}_h}c_{\mathrm{out},i}\left(z_{\mathrm{Rx}_h},t\right)\,.
\end{equation}
 The latter approximation is valid under the \ac{UCA}~\cite{Jamali2019}.

\renewcommand{\arraystretch}{1.35}
\begin{table*}
  \centering
  \caption{Summary of the most important variables, parameters, and operators of the \ac{MIGHT} model.}\label{tab:Notation}
  \resizebox{\textwidth}{!}{%
  \begin{tabular}{|c||c|l|c|c|}
    \hline
    \rowcolor{gray!20}
    &\textbf{Variable/Parameter/Operator} & \textbf{Meaning} & \textbf{Unit} & \textbf{Definition} \\ \hline\hline
    & $t$ & Time & $\SI{}{\second}$ & Sec.~\ref{ssec:SinglePipeModel}\\ \cline{2-5}
    & $D$ & Molecular diffusion coefficient & $\SI{}{\meter\squared\per\second}$ & Sec.~\ref{ssec:DiffusiveMoleculeTransport} \\ \cline{2-5}
    & $\mu$ & Dynamic fluid viscosity & $\SI{}{\kilo\gram\per\meter\per\second}$ & Eq.~\eqref{eqn:HydraulicResistance} \\ \hline\hline
    \multirow{16}{*}{\rotatebox[origin=c]{90}{\textbf{Pipes}}}
    & $E$ & Number of pipes & & Sec.~\ref{ssec:VN_Definition}\\ \cline{2-5}
    & $p_i$ & $i$-th pipe & & Sec.~\ref{ssec:VN_Definition} \\ \cline{2-5}
    & $r_i$ & Radius of pipe $p_i$ & $\SI{}{\meter}$ & Sec.~\ref{ssec:VN_Definition} \\ \cline{2-5}
    & $l_i$ & Length of pipe $p_i$ & $\SI{}{\meter}$ & Sec.~\ref{ssec:VN_Definition} \\ \cline{2-5}
    & $Q_i$ & Flow rate in pipe $p_i$ & $\SI{}{\meter\cubed\per\second}$ & Sec.~\ref{ssec:Equivalent_Circuit_Theory} \\ \cline{2-5}
    & $\bar{u}_i$ & Flow velocity in pipe $p_i$ & $\SI{}{\meter\per\second}$ & Sec.~\ref{ssec:Equivalent_Circuit_Theory} \\ \cline{2-5}
    & $R_i$ & Hydraulic resistance of pipe $p_i$ & $\SI{}{\kilo\gram\per\second\per\meter\tothe{4}}$ & Eq.~\eqref{eqn:HydraulicResistance}\\ \cline{2-5}
    & $\bar{D}_i$ & Effective diffusion coefficient in pipe $p_i$ & $\SI{}{\meter\squared\per\second}$ & Eq.~\eqref{eqn:ArisTaylorEffectiveDiffusionCoefficient} \\ \cline{2-5}
    & $z_i\in [0,l_i]$ & Longitudinal position in pipe $p_i$ & $\SI{}{\meter}$ & Sec.~\ref{ssec:SinglePipeModel} \\ \cline{2-5}
    &$\mu_i(z_i)$ & Mean parameter for pipe molecule flux $j_i(z_i,t)$ & $\SI{}{\second}$ & Eq.~\eqref{eq:pipe-moms} \\ \cline{2-5}
    & $\sigma^2_i(z_i)$ & Variance parameter for pipe molecule flux $j_i(z_i,t)$ & $\SI{}{\second\squared}$ & Eq.~\eqref{eq:pipe-moms} \\ \cline{2-5}
    & $\theta_i(z_i)$ & Shape parameter for pipe molecule flux $j_i(z_i,t)$ & $\SI{}{\second}$ & Eq.~\eqref{eq:pipe-moms} \\ \cline{2-5}
    & $T_i(z_i)$ & \ac{FPT} at position $z_i$ in pipe $p_i$ & $\SI{}{\second}$ & Eq.~\eqref{eq:igd} \\ \cline{2-5}
    & $\mathcal{S}(p_i)$ & Source node of pipe $p_i$ & & Sec.~\ref{ssec:VN_Definition} \\ \cline{2-5}
    & $\mathcal{D}(p_i)$ & Destination node of pipe $p_i$ & & Sec.~\ref{ssec:VN_Definition} \\ \cline{2-5}
    & $j_i(z_i,t)$ & Molecule flux in pipe $p_i$ as response to injection at the pipe inlet  & $\SI{}{\per\second}$ & Eq.~\eqref{eq:pipe-pdf} \\ \hline\hline
    \multirow{10}{*}{\rotatebox[origin=c]{90}{\textbf{Nodes}}}
    & $B$, $I$, $O$, $C$ & Number of bifurcations/inlet nodes/outlet nodes/connecting nodes & & Sec.~\ref{ssec:VN_Definition} \\ \cline{2-5}
    & $n_{\mathrm{in},a}$ & $a$-th inlet node & & Sec.~\ref{ssec:VN_Definition} \\ \cline{2-5}
    & $n_{\mathrm{out},b}$ & $b$-th outlet node & & Sec.~\ref{ssec:VN_Definition}\\ \cline{2-5}
    & $\mathcal{N}_\mathrm{in}$ & Set of inlet nodes & & Sec.~\ref{ssec:VN_Definition} \\ \cline{2-5}
    & $\mathcal{N}_\mathrm{out}$ & Set of outlet nodes & & Sec.~\ref{ssec:VN_Definition} \\ \cline{2-5}
    & $n_{s}$ & $s$-th connecting node & & Sec.~\ref{ssec:VN_Definition} \\ \cline{2-5}
    & $b_m$ & $m$-th bifurcation & & Sec.~\ref{ssec:VN_Definition} \\ \cline{2-5}
    & $\mathcal{O}(b_m)$ & Outflow pipes of bifurcation $b_m$ & & Sec.~\ref{ssec:VN_Definition} \\ \cline{2-5}
    & $Q_{\mathrm{in},a}$ & Inflow rate at inlet $n_{\mathrm{in},a}$ & $\SI{}{\meter\cubed\per\second}$ & Sec.~\ref{ssec:Equivalent_Circuit_Theory} \\ \cline{2-5}
    & $R_{\mathrm{out},b}$ & Terminal resistance at outlet node $n_{\mathrm{out},b}$ & $\SI{}{\kilo\gram\per\second\per\meter\tothe{4}}$ & Sec.~\ref{ssec:Equivalent_Circuit_Theory} \\ \hline\hline
    \multirow{11}{*}{\rotatebox[origin=c]{90}{\textbf{Paths}}}
    & $P_k$ & $k$-th path & & Eq.~\eqref{eqn:PathSet} \\ \cline{2-5}
    & $\mathcal{E}_k$ & Index set of pipes in path $P_k$ & & Eq.~\eqref{eqn:PathSet} \\ \cline{2-5}
    & $\mathcal{B}_k$ & Index set of bifurcations in path $P_k$ & & Eq.~\eqref{eqn:PathSet} \\ \cline{2-5}
    & $p_q$, $p_w$ & First and last pipe in path $P_k$ & & Sec.~\ref{ssec:VN_Definition} \\ \cline{2-5}
    & $\mathcal{P}(n_a,n_b)$ & Set of distinct paths from node $n_a$ to node $n_b$ & & Sec.~\ref{ssec:VN_Definition} \\
    \cline{2-5}
    & $\gamma_{P_k}$ & Fraction of molecules propagating through path $P_k$ & & Eq.~\eqref{eqn:PathWeight} \\ \cline{2-5}
    & $\tilde{\mu}_k$, $\hat{\mu}_k$, $\bar{\mu}_k$ & Mean parameter for molecule flux of path $P_k$ (various cases) & $\SI{}{\second}$  & Eqs.~\eqref{eq:pathMom}, \eqref{eq:mom-het-mu} \\ \cline{2-5}
    & $\tilde{\sigma}^2_k$, $\hat{\sigma}^2_k$, $\bar{\sigma}^2_k$ & Variance parameter for molecule flux of path $P_k$ (various cases) & $\SI{}{\second\squared}$ & Eqs.~\eqref{eq:pathMom}, \eqref{eq:mom-het-sigma} \\ \cline{2-5}
    & $\tilde{\theta}_k$, $\hat{\theta}_k$, $\bar{\theta}_k$ & Shape parameter for molecule flux of path $P_k$ (various cases) & $\SI{}{\second}$ & Eqs.~\eqref{eq:pathMom}, \eqref{eq:mom-het-theta} \\ \cline{2-5}
    & $\tilde{T}_k$, $\hat{T}_k$, $\bar{T}_k$ & \ac{FPT} of path $P_k$ (various cases) & $\SI{}{\second}$ & Eqs.~\eqref{eq:sumt}, \eqref{eq:tk-hom}, \eqref{eq:tk-het}   \\ \cline{2-5}
    & $\bar{j}_k(z_q,z_w,t)$, $\tilde{j}_k(z_q,z_w,t)$, $\hat{j}_k(z_q,z_w,t)$  & Molecule flux at position $z_w$ in pipe $p_w$ of path $P_k$ (various cases) & $\SI{}{\per\second}$ & Theorems~\ref{theorem2}, \ref{theorem3} \\ \hline\hline
    \multirow{12}{*}{\rotatebox[origin=c]{90}{\textbf{\acp{Tx} and \acp{Rx}}}}
    & $\mathrm{Tx}_g$ & $g$-th \ac{Tx} & & Sec.~\ref{ssec:Txs_and_Rxs} \\ \cline{2-5}
    & $z_{\mathrm{Tx}_g,i}$ & Longitudinal position of $\mathrm{Tx}_g$ in pipe $p_i$ & $\SI{}{\meter}$ & Sec.~\ref{ssec:Txs_and_Rxs} \\ \cline{2-5}
    & $\mathcal{S}_\mathrm{Tx}(\mathrm{Tx}_g)$ & Source node of the pipe containing $\mathrm{Tx}_g$ &  & Sec.~\ref{ssec:Txs_and_Rxs} \\ \cline{2-5}
    & $N_{g}$ & Number of molecules injected at $\mathrm{Tx}_g$ &  & Sec.~\ref{ssec:Txs_and_Rxs} \\ \cline{2-5}
    & $f_{\mathrm{Tx}_g}(t)$ & Injection function at $\mathrm{Tx}_g$ & $\SI{}{\per\second}$ & Sec.~\ref{ssec:Txs_and_Rxs} \\ \cline{2-5}
    & $\mathcal{N}_\mathrm{Tx}$ & Set of \acp{Tx} &  & Sec.~\ref{ssec:Txs_and_Rxs} \\ \cline{2-5}
    & $\mathrm{Rx}_h$ & $h$-th \ac{Rx} & & Sec.~\ref{ssec:Txs_and_Rxs} \\ \cline{2-5}
    & $z_{\mathrm{Rx}_h,i}$ & Longitudinal center position of $\mathrm{Rx}_h$ in pipe $p_i$ & $\SI{}{\meter}$ & Sec.~\ref{ssec:Txs_and_Rxs} \\ \cline{2-5}
    & $l_{\mathrm{Rx}_h}$ & Longitudinal extent of $\mathrm{Rx}_h$, centered around $z_{\mathrm{Rx}_h,i}$ & $\SI{}{\meter}$ & Sec.~\ref{ssec:Txs_and_Rxs} \\ \cline{2-5}
    & $\mathcal{N}_\mathrm{Rx}$ & Set of \acp{Rx} &  & Sec.~\ref{ssec:Txs_and_Rxs} \\
    \cline{2-5}
    & $\mathcal{D}_\mathrm{Rx}(\mathrm{Rx}_h)$ & Destination node of the pipe containing $\mathrm{Rx}_h$ &  & Sec.~\ref{ssec:Txs_and_Rxs} \\
    \cline{2-5}
    & $h_{\mathrm{Tx}_g,i}(z_i,t)$ & \ac{CIR} between $\mathrm{Tx}_g$ and position $z_i$ in pipe $p_i$ & $\SI{}{\per\second}$ & Eq.~\eqref{eqn:CIR_between_Tx_and_pi} \\
    \cline{2-5}
    & $j_{\mathrm{out},i}(z_i,t)$, $c_{\mathrm{out},i}(z_i,t)$ & Molecule network flux and concentration at position $z_i$ in pipe $p_i$ & $\SI{}{\per\second}$, $\SI{}{\per\meter}$ & Eqs.~\eqref{eqn:network_flux_arbitrary_inj_functions}, \eqref{eqn:Network_Concentration} \\ \cline{2-5}
    & $N_{\mathrm{Rx}_h}(t)$ & Number of observed molecules within $\mathrm{Rx}_h$ & & Eq.~\eqref{eqn:Nobs} \\ \hline
  \end{tabular}
  }
\end{table*}
\renewcommand{\arraystretch}{1.0}

\subsection{Model Validation}\label{ssec:Model_Validation}

\begin{figure*}
    \centering
    \includegraphics[width=\linewidth]{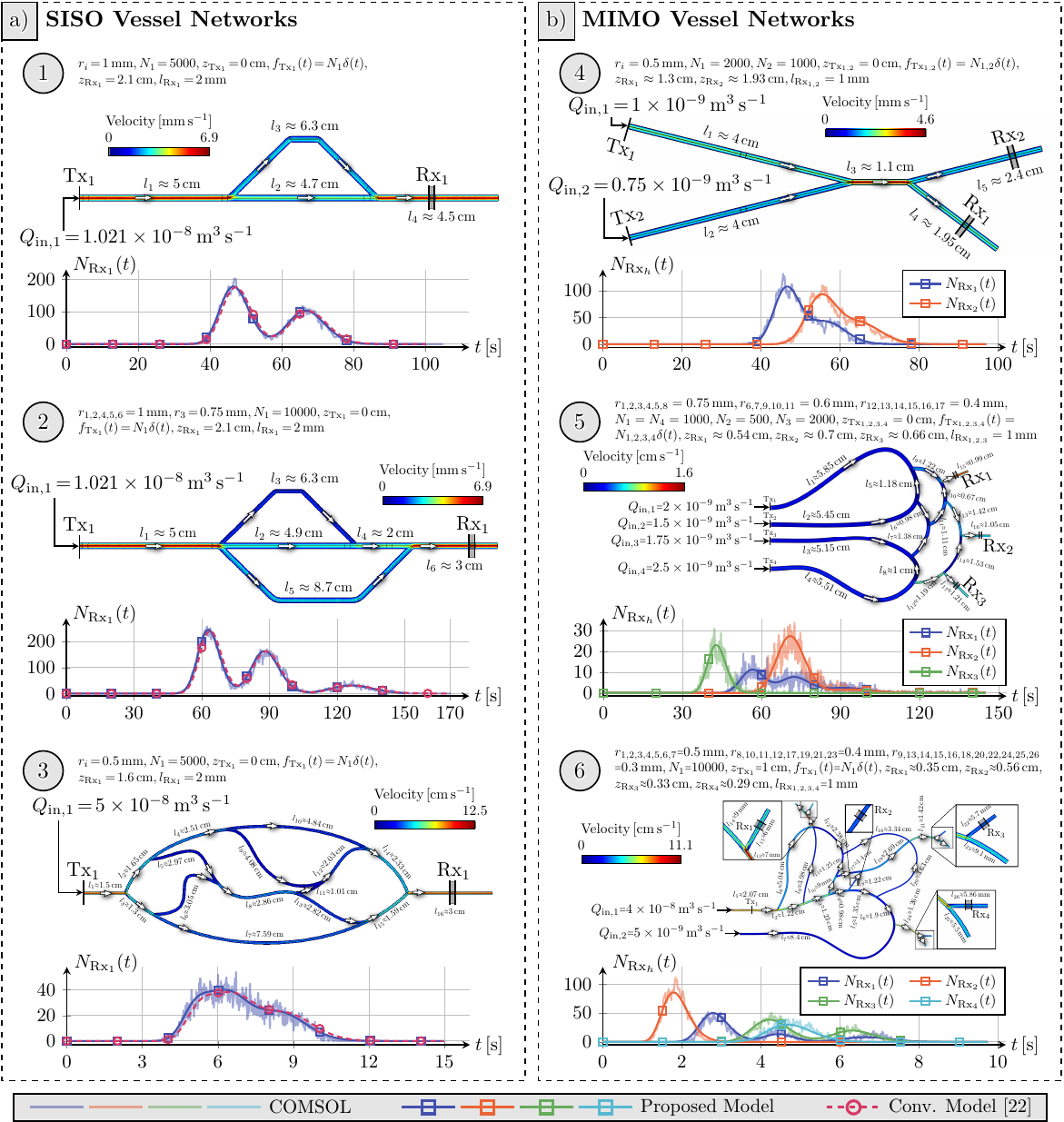}
    \caption{\textbf{Model validation.} 
    The number of molecules $N_{\mathrm{Rx}_h}(t)$ observed at $\mathrm{Rx}_h$ as predicted by~\eqref{eqn:Nobs} is compared to \ac{3-D} COMSOL simulations. COMSOL geometries are shown with color-coded flow velocities, as indicated by the colorbars. Fluid flow directions (white arrows), pipe lengths $l_i$, inflow rates $Q_{\mathrm{in},a}$, and \ac{Tx} and \ac{Rx} positions are annotated. All other relevant parameters are given above the COMSOL geometries, and $D=\SI{1.461e-7}{\meter\squared\per\second}$ holds for all scenarios, motivated by experimental observations in~\cite{Vakilipoor2025}. 
    \textbf{a)}~\ac{SISO}-\acp{VN} (single flow inlet and outlet, single \ac{Tx} and \ac{Rx}) of increasing structural complexity and their received signals. Here, results for the convolution-based model from~\cite{Jakumeit2025} are additionally shown. 
    \textbf{b)}~\ac{MIMO}-\acp{VN} (multiple flow inlets and outlets and/or multiple \acp{Tx} and \acp{Rx}) of increasing structural complexity and their received signals.}
    \label{fig:Model_Validation}
\end{figure*}
To validate the proposed \ac{MIGHT} model, in Fig.~\ref{fig:Model_Validation}, we compare the predicted received number of molecules $N_{\mathrm{Rx}_h}(t)$ in \eqref{eqn:Nobs} for various \ac{SISO}- and \ac{MIMO}-\ac{VN} topologies to the results from finite-element simulations performed in COMSOL. 
COMSOL simulations were carried out in \ac{3-D} space, using the \textit{Laminar Flow} and \textit{Particle Tracing for Fluid Flow} modules with Brownian motion and drag force. 
The inclusion of Brownian motion results in stochastic particle trajectories and, consequently, in noisy and fluctuating received signals.
\textit{Fully Developed Flow}, \textit{No Slip}, and zero static pressure with suppressed backflow were assumed as the flow boundary conditions at the \ac{VN} inlets, pipe walls, and outlets, respectively. 
Particles in COMSOL were initially randomly distributed according to a uniform distribution in the cross-section at the \ac{Tx}(s) position(s). 
As particle boundary condition at the \ac{VN} outlets, \textit{disappear} was selected.
A physics-controlled mesh at \textit{normal} element size was used.

In line with open science principles, all COMSOL simulations presented in this work are openly accessible via Zenodo under the CC BY 4.0 license at \url{https://doi.org/10.5281/zenodo.18461818}. Researchers using this dataset are kindly asked to cite it using the associated Zenodo \ac{DOI}~\cite{jakumeit2026zenodo}.

Fig.~\ref{fig:Model_Validation}a) compares the proposed model in~\eqref{eqn:Nobs}, COMSOL simulations, and the convolution-based model from~\cite{Jakumeit2025} for three \ac{SISO}-\acp{VN} (single flow inlet and outlet, single \ac{Tx} and \ac{Rx}) of increasing structural complexity. All relevant parameters are provided in the figure and its caption.
Across all networks (\ac{VN} 1--3), \ac{MIGHT} shows excellent agreement with the results from COMSOL. 
Notably, it accurately captures the transport dynamics across a wide range of flow velocities, spanning more than one order of magnitude ($\SI{6.9}{\milli\meter\per\second}$ vs.~$\SI{12.5}{\centi\meter\per\second}$), as demonstrated in \acp{VN} 1 and 2 compared to \ac{VN} 3.
While the convolution-based model from~\cite{Jakumeit2025} performs well for the simpler networks (\acp{VN} 1 and 2), it struggles with the more complex topology of \ac{VN} 3, particularly in capturing the signal decay. This limitation arises from numerical inaccuracies introduced by repeated numerical convolutions, where errors accumulate as the number of pipes, and thus the number of convolutions, increases.

Fig.~\ref{fig:Model_Validation}b) shows the comparison between~\eqref{eqn:Nobs} and COMSOL simulations for \ac{MIMO}-\acp{VN}, i.e., networks with multiple flow inlets and outlets, as well as multiple \acp{Tx} and/or \acp{Rx}.
No comparison to the model in~\cite{Jakumeit2025} is made, as it is not applicable for \ac{MIMO}-\acp{VN}.
As before, \acp{VN} 4--6 exhibit increasing structural complexity.
\ac{VN}~4 is the structurally simplest \ac{MIMO}-\ac{VN}, yet the received signals $N_{\mathrm{Rx}_h}(t)$ exhibit non-trivial dynamics. This is due to the asymmetric topology, different inflow rates $Q_{\mathrm{in},1}$ and $Q_{\mathrm{in},2}$, and cross-talk between the \acp{Tx}, all of which are accurately captured by \ac{MIGHT}.
\ac{VN}~5 is inspired by the mesenteric vasculature in the human body, supplying the small intestine. In this scenario, four \acp{Tx} and three \acp{Rx} are simulated in COMSOL. The proposed model closely predicts the resulting received signals. Notably, a comparison of $N_{\mathrm{Rx}_2}(t)$ and $N_{\mathrm{Rx}_3}(t)$ reveals that the earliest arriving signal does not necessarily carry the largest number of molecules. In complex \ac{MIMO}-\acp{VN}, received signal shapes are governed by a combination of factors, including the number and positions of \acp{Tx}, their respective numbers of released molecules, the network topology, and the \acp{Rx}' positions.
\ac{VN}~6 illustrates a broadcasting scenario in which the molecules released by a single \ac{Tx} are observed by multiple \acp{Rx}. Again, the proposed model yields accurate predictions of all received signals. It is also worth noting that \ac{VN}~6 features significantly higher flow velocities compared to \ac{VN}~4 ($\SI{11.1}{\centi\meter\per\second}$ vs.~$\SI{4.6}{\milli\meter\per\second}$).

Overall, our analysis confirms the validity of the \ac{MIGHT} model for describing advection-diffusion-driven hemodynamic molecule transport in complex (\ac{MIMO}) \acp{VN} in the dispersive regime. Across all validation scenarios, for all paths $P_k$, $\mathrm{Pe}_k$ is moderately large and ranges from $20.3$ to $438.0$, rendering the skewness error $\Delta\xi$ in~\eqref{eq:app-err} negligible and further confirming the validity of Theorem~\ref{theorem3}.

\section{Selected Applications of Proposed Model}\label{sec:NumericalResults}
We demonstrate the \ac{MIGHT} model’s utility for three application scenarios: The structural reduction of \ac{SISO}-\acp{VN}, a \ac{Rx}-centric vessel importance scoring in \ac{MIMO}-\acp{VN}, and the estimation of representative \ac{SISO}-\acp{VN} from observed molecular signals. 
These applications are illustrated for \acp{VN} of varying complexity, chosen as physiologically plausible but anatomically abstract test cases. 

\subsection{SISO Vessel Network Reduction}\label{ssec:Vessel_Network_Reduction}
In most \acp{VN}, only certain parts of the topology contribute significantly to the received signal of a given \ac{Rx}, while other parts have negligible influence. Thus, we propose an approach for the systematic reduction of a \ac{SISO}-\ac{VN} (single flow inlet, single flow outlet, single \ac{Tx}, single \ac{Rx}) to a simplified representation that preserves the essential received signal properties at the \ac{Rx}. 
This reduction can facilitate the interpretation of transport dynamics, lower computational cost in large-scale simulations, may lead to surrogate models that capture the behavior of entire organs or tissues without requiring vessel-level detail, and enable efficient parameter estimation or the design of microfluidic testbeds that reproduce dominant transport characteristics of complex \acp{VN}. In the following, we derive a reduction algorithm that identifies those pipes in the network that least significantly contribute to the overall received signal and discards them.

To derive a reduced \ac{VN}, we first compute the \textit{pipe fraction} of each pipe $p_i$, i.e., the fraction of molecules released at $\mathrm{Tx}_g$, propagating through $p_i$, and reaching $\mathrm{Rx}_h$, which is obtained as
\begin{equation}\label{eqn:pipe_fraction}
    \gamma_{\mathrm{Tx}_g,p_i,\mathrm{Rx}_h} = \hspace{-3mm}\sum_{\{ P_k\in\mathcal{P}(\mathcal{S}_\mathrm{Tx}(\mathrm{Tx}_g),\mathcal{D}_\mathrm{Rx}(\mathrm{Rx}_h)) \mid p_i \in P_k \}} \hspace{-3mm}\gamma_{P_k}\,,
    \end{equation}
with $\gamma_{\mathrm{Tx}_g,p_i,\mathrm{Rx}_h}\in[0,1]$, i.e., the fraction of molecules traveling through $p_i$ is the sum of all path fractions of the paths that contain $p_i$. In the \ac{SISO}-\ac{VN} case, we have $\gamma_{\mathrm{Tx}_1,p_i,\mathrm{Rx}_1}$. All pipe fractions $\gamma_{\mathrm{Tx}_1,p_i,\mathrm{Rx}_1}$ and associated pipes are then sorted in descending order, and pipes whose pipe fractions lie in the lower $\alpha$-quantile are discarded from the \ac{SISO}-\ac{VN} topology, retaining only the most significant parts of the network. The parameter $\alpha \in [0,1]$ is a pruning parameter and controls the degree of reduction.
\begin{figure}
    \centering
    \includegraphics[width=\linewidth]{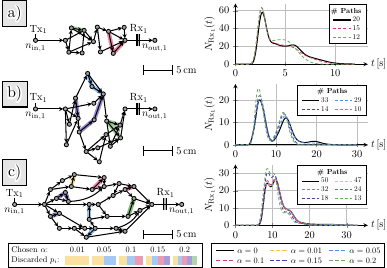}
    \caption{\textbf{Structural reduction of \ac{SISO}-\acp{VN}.} For each \ac{VN}, the original topology and the discarded pipes, as well as the resulting $N_{\mathrm{Rx}_1}(t)$ are shown for various $\alpha$. Pipe lengths are depicted to scale. In \textbf{a)} and \textbf{b)}, radii were drawn from a normal distribution with mean $\SI{0.5}{\milli\meter}$ and standard deviation $\SI{0.1}{\milli\meter}$. In \textbf{c)}, $r_i=\SI{0.5}{\milli\meter}$.}
    \label{fig:Vessel_Network_Reduction}
\end{figure}

The \ac{VN} reduction is illustrated in Fig.~\ref{fig:Vessel_Network_Reduction} for various \ac{SISO}-\acp{VN} and $\alpha$-values, with $N_1=10000$, $Q_{\mathrm{in},1}=\SI{1e-7}{\meter\cubed\per\second}$, $D=\SI{1.461e-7}{\meter\squared\per\second}$, $z_{\mathrm{Tx}_1}=\SI{0}{\meter}$, $z_{\mathrm{Rx}_1}=\SI{2.1}{\centi\meter}$, and $l_{\mathrm{Rx}_1}=\SI{2}{\milli\meter}$. 
We observe that larger $\alpha$ result in more aggressive pruning. 
For all \acp{VN}, appropriate $\alpha$ substantially reduce the structural \ac{VN} complexity while preserving the overall transport dynamics. 
The retention of the transport dynamics can be observed from the received signals, that retain their characteristic shapes, even after the \ac{VN} reduction.
For instance, in Fig.~\ref{fig:Vessel_Network_Reduction}c), the received signals for $\alpha=\{0.01,0.05,0.1 \}$ (yellow, blue, and rose curves) closely resemble the original received signal. Only once $\alpha$ gets larger ($\alpha=\{ 0.15,0.2\}$, purple and green curves), the signals of the reduced \acp{VN} start to significantly deviate from the original signal. 
The number of paths between $n_{\mathrm{in},1}$ and $n_{\mathrm{out},1}$ can often be halved without markedly affecting molecule transport. Since the removal of a single pipe potentially eliminates multiple paths, a far stronger reduction in the number of paths than in the number of vessels is generally possible.
Note that in reduced \acp{VN}, the received signal may temporarily exceed that of the original network. This results from mass conservation: when certain pipes and their associated paths are removed, molecules are redistributed over the rest of the network, increasing flow and concentration in the remaining paths. Consequently, some time intervals exhibit amplified signals, as seen, e.g., in Fig.~\ref{fig:Vessel_Network_Reduction}a), where the signal of the 12-path reduced \ac{VN} (green curve) partially exceeds that of the original \ac{VN} (black curve).

\subsection{Vessel Importance Scoring in MIMO Vessel Networks}

\begin{figure*}
    \centering
    \includegraphics[width=\linewidth]{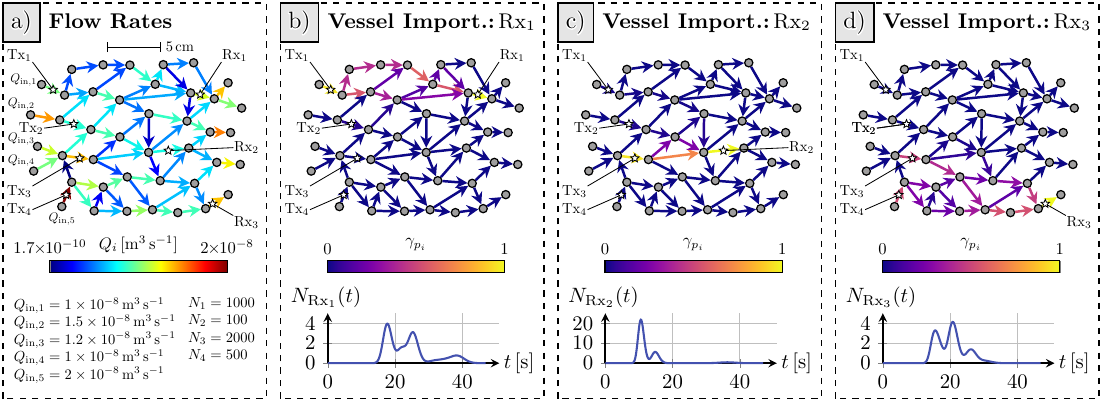}
    \caption{\textbf{Vessel importance scoring in \ac{MIMO}-\acp{VN}.} \textbf{a) Flow rates:} The flow rates for an exemplary densely connected \ac{MIMO}-\ac{VN} are color-coded in the edge colors. Additionally, the positions of four \acp{Tx} and three \acp{Rx} are indicated, as well as the applied flow rates at the inlets and the number of released molecules at each \ac{Tx}. \textbf{b)-d) Vessel importance:} The spatial distributions of the vessel importance scores in~\eqref{eqn:pipe_frac_times_Tx_impact} are color-coded for three different \acp{Rx}. Received signals are shown alongside the \acp{VN}.}
    \label{fig:vessel_importance}
\end{figure*}

Similar to \ac{SISO}-\acp{VN}, for a given \ac{Rx} in a complex, large-scale \ac{MIMO}-\ac{VN}, different regions of the network contribute unequally to the received signal, as they transport different amounts of signaling molecules. 
However, in \ac{MIMO}-\acp{VN}, the importance of individual vessels is shaped by a variety of interacting factors, including the positions of the \acp{Tx} and the numbers of molecules they release, the applied flow rates at the inlets, the network topology, the placement of the \acp{Rx} and flow outlets, and the terminal resistances at the outlets.
Moreover, signaling molecules may be "lost" within the network and never arrive at the considered \ac{Rx}.
These factors jointly induce complex, non-intuitive transport dynamics, making it generally infeasible to assess vessel importance through simple reasoning alone.

Nevertheless, quantifying vessel importance can be crucial for practical tasks: It enables targeted sensor placement in high-signal regions, but also helps identify low-impact regions, which is useful, e.g., when minimizing sensor interference or selecting robust sites for signal-insensitive implants. 
The analytical tractability of the \ac{MIGHT} model enables the derivation of metrics that quantify the role of each vessel in molecule transport, even in arbitrary \ac{MIMO}-\acp{VN}. 
In the following, we propose such a metric and apply it to a large-scale \ac{MIMO}-\ac{VN} to demonstrate its practical utility.

Consider $\mathrm{Rx}_h$, positioned in pipe $p_\nu$ in a given \ac{MIMO}-\ac{VN}.
Additionally to the \textit{pipe fraction} metric in~\eqref{eqn:pipe_fraction}, we propose the \textit{\ac{Tx} impact} metric. In general, multiple \acp{Tx}, denoted by $\mathrm{Tx}_g\in\mathcal{N}_\mathrm{Tx}$, contribute to the received signal $N_{\mathrm{Rx}_h}(t)$. The relative importance of the $e$-th \ac{Tx}, denoted by $\mathrm{Tx}_e$, can then be expressed as a simple fraction
    \begin{equation}\label{eqn:Tx_impact}
        \gamma_{\mathrm{Tx}_e} = \frac{N_e}{\sum_{N_h\in\mathcal{N}_\mathrm{Tx}}N_h},
    \end{equation}
with $\gamma_{\mathrm{Tx}_e}\in[0,1]$, i.e., the \textit{\ac{Tx} impact} depends on the fraction of signaling molecules that were released by the considered \ac{Tx}, compared to the total number of signaling molecules released by all \acp{Tx}. 
Importantly, the metrics in~\eqref{eqn:pipe_fraction} and~\eqref{eqn:Tx_impact} reflect the long-term cumulative number of molecules being transported through any given pipe and is thus invariant to the temporal profiles of the injection functions $f_{\mathrm{Tx}_g}(t), \mathrm{Tx}_g\in\mathcal{N}_{\mathrm{Tx}}$. Only the numbers of released molecules, $N_g$, play a role for the metrics.

Overall, the importance score of pipe $p_i$ for received signal $N_{\mathrm{Rx}_h}(t)$ is then obtained as
\begin{equation}\label{eqn:pipe_frac_times_Tx_impact}
    \gamma_{p_i}=\frac{1}{\Psi}\sum_{\mathrm{Tx}_g\in\mathcal{N}_\mathrm{Tx}} \underbrace{\gamma_{\mathrm{Tx}_g}}_{\text{Tx impact}}\hspace{2mm}\underbrace{\gamma_{\mathrm{Tx}_g,p_i,\mathrm{Rx}_h}}_{\text{Pipe fraction}}\,,
\end{equation}
with $\gamma_{p_i}\in [0,1]$.
Here,
\begin{equation}
    \Psi = \sum_{\mathrm{Tx}_g\in\mathcal{N}_\mathrm{Tx}}\gamma_{\mathrm{Tx}_g,p_\nu,\mathrm{Rx}_h},
\end{equation}
with $\Psi\in [0,1]$,
is the total fraction of molecules arriving at $\mathrm{Rx}_h$. 
The metric in~\eqref{eqn:pipe_frac_times_Tx_impact} is thus computed by summing over all \acp{Tx}, weighting each \ac{Tx}'s impact by the fraction of molecules traversing the considered pipe along all paths from that \ac{Tx} to $\mathrm{Rx}_h$, and by normalizing the result by the total fraction of molecules arriving at $\mathrm{Rx}_h$. By design, this ensures that $\gamma_{p_\nu} = 1$ for the pipe $p_\nu$ containing the considered \ac{Rx}, assigning it the highest possible importance score. Conversely, any pipe $p_i$ that does not lie on any path from any \ac{Tx} to $\mathrm{Rx}_h$ yields $\gamma_{p_i} = 0$.

The vessel importance score is evaluated for an exemplary \ac{MIMO}-\ac{VN} in Fig.~\ref{fig:vessel_importance}. The graph representation of the \ac{VN} is shown in Fig.~\ref{fig:vessel_importance}a), with vessel-wise color-coded flow rates. The network includes five flow inlets with different applied flow rates, five flow outlets, four \acp{Tx} with varying molecule release quantities and $f_{\mathrm{Tx}_g}(t)=N_g\delta (t), \forall \mathrm{Tx_g\in\mathcal{N}_\mathrm{Tx}}$, and three \acp{Rx}, each modeled as having a length of $l_{\mathrm{Rx}_h} = \SI{1}{\milli\meter}$. All pipe lengths are drawn to scale and share a uniform radius of $r_i = \SI{0.5}{\milli\meter}$. In total, the \ac{VN} contains more than 400 distinct paths connecting all inlet-outlet pairs.

Since the \ac{VN} is densely connected and exhibits a relatively uniform structure, the flow distribution in Fig.~\ref{fig:vessel_importance}a) shows that the largest flow rates occur at the inlets and outlets. In contrast, the flow rates in the central region of the network are lower, as the flow splits among the branches according to the law of mass conservation.
Figs.~\ref{fig:vessel_importance}b)--d) highlight the spatial distribution of vessel importance scores for the different \acp{Rx} in the \ac{VN}. As expected, the vessels located upstream of each \ac{Rx}, particularly those forming a cone-shaped region feeding into the \ac{Rx}, exhibit the highest importance. Conversely, vessels that are downstream or distant from the \ac{Rx} tend to contribute less to molecule transport and thus receive lower importance scores.
Notably, the vessels that contain the \acp{Rx} are consistently the most important ones, with $\gamma_{p_\nu}=1$. More generally, for vessels located in the upstream region of a given \ac{Rx}, higher flow rates roughly correlate with higher importance scores. This reflects the physical tendency of molecules to follow the path of least resistance, which is governed by the flow distribution.
The influence of the number of molecules released at the \acp{Tx} is also evident. For instance, in Fig.~\ref{fig:vessel_importance}c), vessels originating from $\mathrm{Tx}_2$ exhibit low importance despite their proximity to $\mathrm{Rx}_2$. This results from $\mathrm{Tx}_2$ releasing a substantially smaller number of molecules ($N_2=100$) compared to the other \acp{Tx} (e.g., $N_3=2000$) and demonstrates how low source strength can diminish the downstream vessel importance, even under favorable geometric conditions.
Overall, these results demonstrate that the proposed vessel importance metric in~\eqref{eqn:pipe_frac_times_Tx_impact} provides an effective and intuitive means of quantifying the importance of different regions within a \ac{VN}, enabling rapid insight into the complex transport behavior of large-scale \ac{MIMO}-\acp{VN}.

Note that \ac{VN} reduction based on~\eqref{eqn:pipe_frac_times_Tx_impact}, as illustrated for \ac{SISO}-\acp{VN} in Subsection~\ref{ssec:Vessel_Network_Reduction}, is generally not applicable to \ac{MIMO}-\acp{VN}. This is because pruning vessels solely based on their importance scores can yield invalid topologies, e.g., disconnected path segments that are no longer connected to any \ac{Tx}. In contrast to the \ac{SISO} case, the presence of multiple \acp{Tx}, \acp{Rx}, and flow inlets/outlets in \ac{MIMO} networks introduces dependencies that must be respected during reduction. Therefore, more sophisticated algorithms are required that incorporate structural clean-up steps to ensure topological consistency. The development of such algorithms is left for future work.

\subsection{Estimating Vessel Networks from Signals}

Estimating a representative \ac{VN} from an observed molecular signal is crucial for many envisioned \ac{MC} applications, especially since patient-specific \ac{VN} topologies are generally unknown. For instance, estimating \acp{VN} could yield simplified \acp{VN} with comparable transport dynamics, useful for designing microfluidic chips that mimic organ dynamics. In the following, leveraging the tractability of the \ac{MIGHT} model, we propose a method for estimating a simplified \ac{SISO}-\ac{VN} that reproduces the behavior of an unknown complex \ac{SISO}-\ac{VN} underlying an observed signal $r(t)$.
For brevity, we focus on the \ac{SISO} case with impulsive injections at each \ac{Tx}, since the focus of this work is the introduction of \ac{MIGHT} and not its applications. In future work, we aim to generalize the approach to \ac{MIMO}-\acp{VN}. 

\begin{figure}
    \centering
    \includegraphics[width=\linewidth]{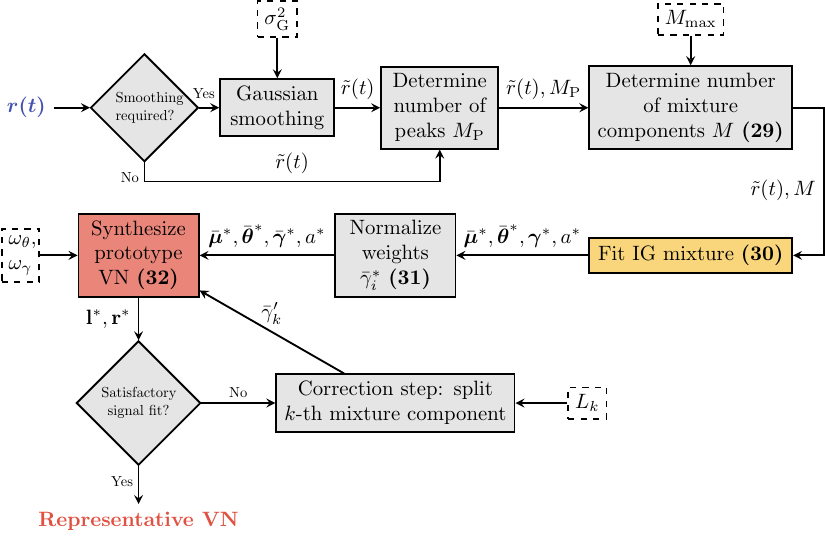}
    \caption{\textbf{VN estimation procedure.} 
    Beginning with a molecular signal $r(t)$, the estimation procedure outputs a representative \ac{VN} that closely replicates the signal. The core steps, i.e., fitting the \ac{IG} mixture and synthesis of a prototype \ac{VN}, are illustrated in yellow and red, respectively. Hyper-parameters are indicated by dashed boxes and respective equations are given inside the solid boxes.}
    \label{fig:block_diagram_VN_estimation}
\end{figure}

Our approach for \ac{VN} estimation comprises two main stages: First, we fit a mixture of \ac{IG} distributions to the observed molecular signal. Then, we iteratively synthesize a prototype \ac{VN} capable of reproducing the estimated mixture from the first step. 
The individual steps of the \ac{VN} estimation procedure are summarized in~Fig.~\ref{fig:block_diagram_VN_estimation}.

\begin{figure*}
    \centering
    \includegraphics[width=\linewidth]{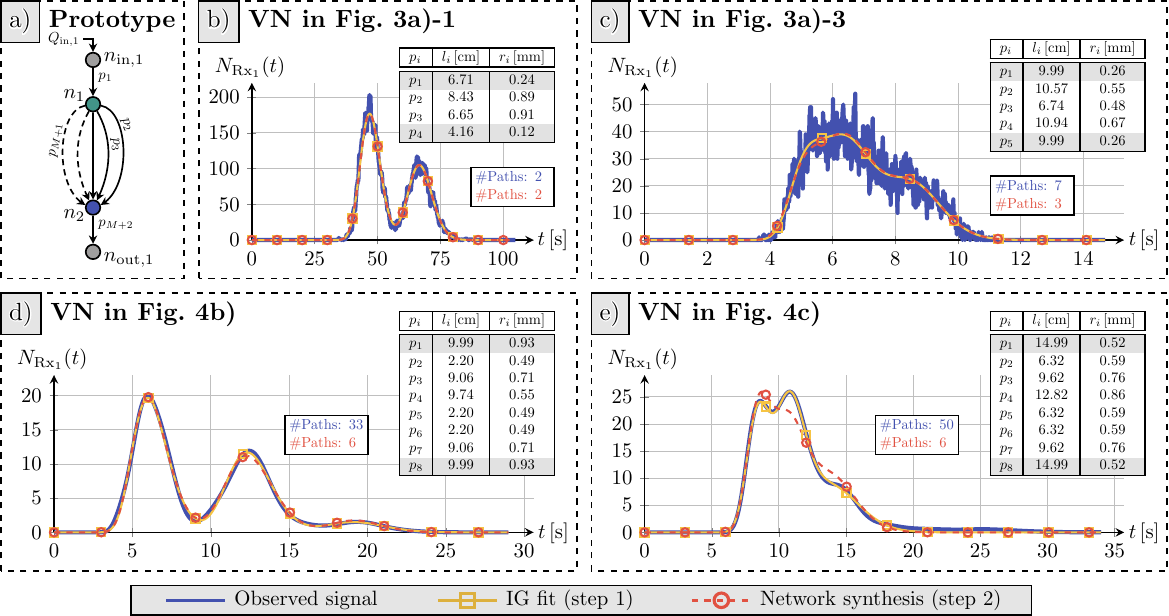}
    \caption{\textbf{\ac{VN} estimation results.} \textbf{a)} Structure of prototype \ac{VN}. \textbf{b)--e)} Estimated \acp{VN} from COMSOL data (b), c)) and synthetic signals (d), e)) reproduce the observed signals across different \ac{VN} sizes. Fitted \ac{IG} signals and estimated outputs are shown in yellow and red, respectively. The numbers of paths in the original and estimated \acp{VN} are shown in dark blue and red in the subplots. Inlet and outlet pipes are highlighted in light gray.}
    \label{fig:Network_Estimation}
\end{figure*}

The \textbf{fitting procedure} employs as input an observed molecular signal $r(t)$, e.g., $N_{\mathrm{Rx}_1}(t)$, flux, or concentration, and the maximum number of fitting components $M_{\max}$ and returns the mean, scale, and weight parameters of the fitted mixture with $M$ components. 
In a pre-processing step, we smooth $r(t)$, if it is a noisy particle count (i.e., $r(t) = N_{\mathrm{Rx}}(t)$ as obtained from COMSOL simulations), with a Gaussian filter with a standard deviation of $\sigma_{\mathrm{G}}=10$ samples to obtain $\Tilde{r}(t)$. We determine the number of peaks $M_{\mathrm{P}}$ in the obtained signal using SciPy's \verb|find_peaks| function. Then, we choose the number of mixture components $M$ according to
\begin{equation}
    M = \min\{M_{\max}, M_{\mathrm{P}}\}.
\end{equation}
Finally, we optimize the parameters $\bar{\boldsymbol{\mu}}=[\bar{\mu}_1, \dots, \bar{\mu}_M]^{\mathrm{T}}$, $\bar{\boldsymbol{\theta}}=[\bar{\theta}_1, \dots, \bar{\theta}_M]^{\mathrm{T}}$, and $\boldsymbol{\gamma}=[\gamma_{1}, \dots, \gamma_{M}]^{\mathrm{T}}$ of the resulting \ac{IG} mixture by solving the following optimization problem: 
\begin{equation}\label{eqn:Optimal_Mean_and_Scale}
\begin{aligned}
    \{\bar{\boldsymbol{\mu}}^*\!,\bar{\boldsymbol{\theta}}^*\!\!, \boldsymbol{\gamma}^*\!\!,a^*\}\!\!&=\!\argmin_{\bar{\boldsymbol{\mu}},
        \bar{\boldsymbol{\theta}}, \boldsymbol{\gamma},a}
    \!\int\limits_{0}^{\infty}\!\!
    \left(
        \!\Tilde{r}(t) \!-\! a\!\!\sum_{k=1}^{M}\!\! \gamma_{k} \bar{j}_k\!\left(l_w, t;0\right)\big|_{\bar{\mu}_k,\bar{\theta}_k}\!\!
    \right)^{\hspace*{-1mm}2}\hspace*{-1mm}\mathrm{d}t\\
    &\mathrm{s.t.}\;\; \forall k: \gamma_{k}\geq0\;,\; \bar{\mu}_k\geq0,\; \bar{\theta}_k\geq0\,.
\end{aligned}
\end{equation} 
Here, $a$ is a fitted scaling factor that may be needed to translate the fitted flux signal to a concentration or $N_{\mathrm{Rx}_1}(t)$, see~\eqref{eqn:Network_Concentration} and \eqref{eqn:Nobs}.
As can be seen from~\eqref{eqn:Optimal_Mean_and_Scale}, the fitting procedure does not enforce that $\sum_{i=1}^{M} \gamma_M = 1$ as would be required for the network flux in a \ac{SISO}-\ac{VN}. Therefore, in the remainder of the algorithm, we use normalized weights\footnote{Normalizing the weights only \textit{after} the fitting procedure, rather than enforcing normalization during fitting, facilitates the identification of a well‑fitting mixture, while obeying physical constraints.} $\bar{\gamma}_k^*$, obtained as:
\begin{equation}
    \bar{\gamma}_i^* = \frac{\gamma_i^*}{\sum_{j=1}^{M} \gamma_j^*}\,.
\end{equation}

Next, we \textbf{synthesize a prototype \ac{VN}} with $M$ parallel paths as depicted in \mbox{Fig.~\ref{fig:Network_Estimation}a)}.
This prototype \ac{VN} topology is chosen because it is the simplest topology that exhibits parallel paths, which can be directly mapped to the corresponding individual sum terms in the \ac{MIGHT} model.
We collect the prototype's pipe radii and lengths in vectors $\mathbf{r} = [r_1, \dots, r_E]$ and $\mathbf{l} = [l_1, \dots, l_E]$, respectively, with $E=M+2$. 
Initially, we fit $\mathbf{r}$ and $\mathbf{l}$ so that $\bar{\boldsymbol{\mu}}^*\!,\bar{\boldsymbol{\theta}}^*\!\!, \bar{\boldsymbol{\gamma}}^*$ are closely resembled in a weighted least-squares sense using SciPy's Trust Region Reflective algorithm according to:
\begin{equation}\label{eq:vn_synthesis}
    \begin{aligned}
        \mathbf{l}^*\!\!, \mathbf{r}^* &\!=\argmin_{\mathbf{l}, \mathbf{r}}
         \lVert  \boldsymbol{\bar{\mu}}^*\!\!- \boldsymbol{\bar{\mu}} \rVert_2^2 + w_\theta \lVert \boldsymbol{\bar{\theta}}^*\!\!-\boldsymbol{\bar{\theta}} \rVert_2^2 + w_\gamma \lVert \boldsymbol{\bar{\gamma}}^*\!\! - \boldsymbol{\bar{\gamma}} \rVert_2^2 \\
        &\hspace*{-4mm}\mathrm{s.t.}\hspace{1mm}\bar{u}_1 \hspace{-0.8mm}=\hspace{-0.8mm} \frac{Q_{\mathrm{in},1}}{\pi r_1^2}, \hspace{1mm} \bar{u}_E \hspace{-0.8mm}=\hspace{-0.8mm} \frac{Q_{\mathrm{in},1}}{\pi r_E^2}, 
        \\
        &\hspace*{-5mm}\forall k\!: \hspace{1mm} \bar{u}_{k+1} \hspace{-0.8mm}=\hspace{-0.8mm} \frac{Q_{\mathrm{in},1}}{\pi r_{k+1}^2}  \frac{r_{k+1}^4 / l_{k+1}}{\sum_{\kappa=1}^{M} r_\kappa^4 / l_\kappa},\\
        &\bar{\mu}_k \hspace{-0.8mm}=\hspace{-0.8mm} \mu_1 \hspace{-0.8mm}+\hspace{-0.8mm} \mu_{k+1}\hspace{-0.8mm} +\hspace{-0.8mm} \mu_E,\hspace{1mm}\bar{\sigma}^2_k \hspace{-0.8mm}=\hspace{-0.8mm} \sigma_1^2 \hspace{-0.8mm} + \hspace{-0.8mm}\sigma_{k+1}^2 \hspace{-0.8mm}+\hspace{-0.8mm}\sigma_E^2,\hspace{1mm} \!\bar{\theta}_k \hspace{-0.8mm}=\hspace{-0.8mm}\frac{\bar{\sigma}_k^2}{\bar{\mu}_k}\,.
    \end{aligned}
\end{equation}
Here, $w_\theta, w_\gamma \geq 0$ are hyper-parameters that are chosen such that $\boldsymbol{\bar{\mu}}$, $\boldsymbol{\bar{\theta}}$, and $\boldsymbol{\bar{\gamma}}$ contribute on comparable scales to the objective function. For brevity, we drop the positional arguments $0$ and $l_E$ in \eqref{eq:vn_synthesis}. 

Since \eqref{eq:vn_synthesis} is a non-linear problem and the parameters of the different paths are coupled non-linearly through~\eqref{eqn:PathWeight}, suitable parameters $\mathbf{l}^*$, $\mathbf{r}^*$ may not always be found directly, i.e., the resulting \ac{VN} may not be able to reproduce the dynamics of $\tilde{r}(t)$ sufficiently. 
This can happen because the magnitudes of the weight factors $\boldsymbol{\bar{\gamma}}^*$ differ too much. In such cases, we increase $M$ by splitting the $k$-th mixture component with the largest value of $\bar{\gamma}^*_k$ into $L_k$ components with identical $\bar{\mu}_k^*$ and $\bar{\theta}_k^*$ but reduced weights $\bar{\gamma}'_{k} \leftarrow \bar{\gamma}^*_{k}/L_k$. This step is repeated until a suitable prototype \ac{VN} is identified. 

Fig.~\ref{fig:Network_Estimation} shows several observed signals of four previously introduced \ac{SISO}-\acp{VN} (blue), the signals resulting from the fitted \ac{IG} mixture (yellow), the parameters of the prototype \ac{VN}, and the signal resulting from the prototype \ac{VN} (red). We observe that the proposed approach for \ac{VN} estimation is capable of reliably identifying prototype \acp{VN} that can reproduce a wide variety of signal dynamics with different time scales and shapes. 
In Fig.~\ref{fig:Network_Estimation}b), fitted and estimated signal coincide and the underlying \ac{VN} has the same topology as the prototype \ac{VN} (see Figs.~\ref{fig:Model_Validation}a)-1 and \ref{fig:Network_Estimation}a)). 
In Figs.~\ref{fig:Network_Estimation}c)--e), the estimated networks contain far fewer paths compared to the original ones, demonstrating that a few dominant paths suffice to capture complex \ac{VN} behavior. In Figs.~\ref{fig:Network_Estimation}d) and~\ref{fig:Network_Estimation}e), $p_5$ and $p_6$ take on identical values as their mean and scale parameters coincide due to the correction step enforcing smaller weights. Overall, these results confirm the feasibility of \ac{VN} estimation using the proposed approach.

\section{Conclusion}\label{sec:Conclusion}
In this paper, we proposed a novel model for advective-diffusive \ac{MC} in complex \ac{MIMO}-\acp{VN}, i.e., \acp{VN} comprising multiple flow inlets and outlets, as well as multiple \acp{Tx} and \acp{Rx}.
The model, termed \ac{MIGHT}, is derived from first principles and based on the concept of \ac{FPT}. 
It expresses network-wide molecule transport dynamics in terms of weighted sums of \ac{IG} distributions, parameterized by the physical properties of the \ac{VN} and signaling molecules, and avoids the convolutions present in many existing models. 
As such, \ac{MIGHT} provides a tractable and physically interpretable mathematical framework for \ac{MC} in complex \acp{VN}.
The model accuracy was validated via \ac{3-D} finite-element simulations in COMSOL and by comparison to the convolution-based model in~\cite{Jakumeit2025}, both for \ac{SISO}- and \ac{MIMO}-\acp{VN} of varying structural complexity.
Leveraging the model's analytical tractability, we further proposed methods for the structural reduction of \ac{SISO}-\acp{VN}, the quantification of vessel importance in \ac{MIMO}-\acp{VN}, and the estimation of simplified \ac{SISO}-\acp{VN} that replicate molecular signals from unknown \ac{SISO} topologies.
In summary, the \ac{MIGHT} model combines physical accuracy with analytical tractability and constitutes a practical tool for analyzing, simplifying, and inferring \acp{VN} in \ac{MC}.

Future work will follow two main avenues. First, the current model can be extended to incorporate additional physical phenomena relevant to molecule transport in \acp{VN}. This includes time-varying flow at the inlets, reversible sorption at the vessel walls, gravitational effects, and chemical degradation or absorption of signaling molecules. Furthermore, extending the model to closed-loop \acp{VN} would allow the analysis of complete circulatory systems, including \ac{ISI} and inter-loop-interference effects. On the communications side, the development of optimal pulse shapes, equalization methods, and detection strategies tailored to complex \ac{MIMO}-\acp{VN} are promising directions. Additionally, channel noise models may be incorporated into the existing framework. Moreover, further validation should be pursued using \acp{VN} extracted from clinical imaging data, combined with experimental molecule distribution measurements from \textit{in-vitro} systems (e.g., microfluidic chips) and \textit{in-vivo} models (e.g., rodent vasculature or the chorioallantoic membrane (CAM)~\cite{Vakilipoor2025}). 

Second, \ac{MIGHT} opens the door to a range of \ac{MC} applications. One promising direction is the experimental validation of the proposed estimation scheme for simplified \acp{VN}. This could allow complex anatomical structures to be represented by reduced surrogate networks that preserve key signal characteristics, thereby enabling more tractable experimental setups that mimic specific regions of human vasculature. Additionally, the \ac{MIGHT} model may serve as a foundation for molecule source localization algorithms in \acp{VN}, which are envisioned for diagnostic applications where the point of emission of the signaling molecules must be inferred from sensor data.

\appendix[]

\subsection{Proof of Theorem 1}
\label{sec:proof1}
\begin{proof}
The \ac{1-D} advection-diffusion equation for a pipe $p_i$, where a single molecule enters at $z_i = 0$ at time $t = 0$, is given by 
\begin{align}
    \partial_t c_i(z_i,t) = \bar{D}_i\partial_{z_iz_i}c_i(z_i,t) - \bar{u}_i\partial_{z_i} c_i(z_i,t),
    \label{eq:pde}
\end{align}
with the initial condition $c_i(z_i,0) = 0$ and the influx boundary condition at the pipe inlet 
\begin{align}
    -\bar{D}_i\partial_{z_i} c_i(0,t) + \bar{u}_i c_i(0,t) = \delta(t), 
    \label{eq:bc}
\end{align}
where $\partial_t$ and $\partial_{z_i}$ denote the first partial derivatives \ac{wrt} $t$ and $z_i$, respectively, and $\partial_{z_iz_i}$ denotes the second partial derivative \ac{wrt} $z_i$. Here, without loss of generality, we consider the pipe inlet ($z_i = 0$) as the point of influx. An influx at any other point along the domain of the pipe $z_i\in[0,l_i]$ could be modeled equivalently by shortening the pipe. However, the assumption of influx at $z_i=0$ simplifies the subsequent derivations.

By applying the one-sided Laplace transform \ac{wrt} $t$, denoted by $\mathcal{L}\{ \cdot\}$, and exploiting the initial condition $c_i(z_i,0) = 0$, \eqref{eq:pde} can be transformed into the frequency domain as follows 
\begin{align}
    \bar{D}_i \partial_{z_iz_i}C_i(z_i,s) - \bar{u}_i\partial_{z_i}C(z_i,s) - sC_i(z_i,s) = 0,
    \label{eq:pdeLap}
\end{align}
where $s$ denotes the complex frequency variable of the Laplace transform and $C_i(z_i,s) = \mathcal{L}\{c_i(z_i,t)\}$ denotes the Laplace-transformed concentration.

Next, we assume a standard solution for the concentration $C_i(z_i,s)$ in the Laplace domain of the form
\begin{align}
    C_i(z_i,s) = A_i(s)\mathrm{e}^{\lambda_i(s,\bar{u}_i,\bar{D}_i) z_i},
    \label{eq:solLap}
\end{align}
where $A_i(s)$ and $\lambda_i(s,\bar{u}_i,\bar{D}_i)$ are to be determined.
Inserting \eqref{eq:solLap} into \eqref{eq:pdeLap} and solving for the $\lambda_i$-values leads to
\begin{align}
    \lambda_i(s,\bar{u}_i,\bar{D}_i) = \frac{\bar{u}_i \pm \sqrt{\bar{u}_i^2 + 4\bar{D}_i s}}{2\bar{D}_i}\,.
    \label{eq:lambda}
\end{align}
Since $c_i(z_i,t) \to 0$ for $t\to\infty$, we only consider the $\lambda_i$-value in \eqref{eq:lambda} with the minus sign in the $\pm$-operator. 

To obtain $A_i(s)$, we insert \eqref{eq:solLap} into the Laplace transform of boundary condition \eqref{eq:bc}, leading to 
\begin{align}
    -\bar{D}_i\lambda_iA_i(s) + \bar{u}_iA_i(s) = 1 
    \hspace*{1.5mm}\Leftrightarrow\hspace*{1.5mm} 
    A_i(s) = \frac{1}{\bar{u}_i - \bar{D}_i\lambda_i},
    \label{eq:Ai}
\end{align}
where we dropped the arguments of $\lambda_i$ for better readability.
Inserting \eqref{eq:lambda} and \eqref{eq:Ai} into \eqref{eq:solLap} yields the desired solution for $C_i(z_i,s)$.

Subsequently, we derive a solution for the advective-diffusive flux $j_i(z_i,t)$ in pipe $p_i$, defined as
\begin{align}
    j_i(z_i,t) = \bar{u}_ic_i(z_i,t) - \bar{D}_i\partial_{z_i}c_i(z_i,t)\,.
    \label{eq:flux}
\end{align}
An expression for $J_i(z_i,s) = \mathcal{L}\{j_i(z_i,t)\}$ in the Laplace domain can be obtained by transforming \eqref{eq:flux} into the Laplace domain and inserting $C_i(z_i,s)$ from \eqref{eq:solLap} as follows
\begin{align}
    J_i(z_i,s) &= \bar{u}_iC_i(z_i,s)-\bar{D}_i\partial_{z_i}C_i(z_i,s) \nonumber\\
    &= \underset{A_i(s)^{-1}}{\underbrace{(\bar{u}_i - \lambda_i\bar{D}_i)}} A_i(s)\mathrm{e}^{\lambda_i z_i} = \mathrm{e}^{\lambda_i z_i}\,.
    \label{eq:jilap}
\end{align}
Next, we obtain the time-domain flux $j_i(z_i,t)$ by applying the inverse one-sided Laplace transform $\mathcal{L}^{-1}\{\cdot\}$ to \eqref{eq:jilap}:
\begin{align}
    j_i(z_i,t) = \mathcal{L}^{-1}\{J_i(z_i,s)\} = \frac{z_i}{\sqrt{4\pi \bar{D}_i t^3}}\mathrm{e}^{\left(-\frac{(z_i - \bar{u}_i t)^2}{4\bar{D}_it}\right)}\,.
    \label{eq:pde-flux}
\end{align}
Exploiting the definitions for the mean, variance, and scale parameter from \eqref{eq:pipe-moms}, \eqref{eq:pde-flux} follows as
\begin{align}
    j_i(z_i,t) &= \frac{z_i}{\sqrt{4\pi \bar{D}_i t^3}}\mathrm{e}^{\left(-\frac{(z_i - \bar{u}_i t)^2}{4\bar{D}_it}\right)} = 
    \frac{\frac{z_i}{\bar{u}_i}}{\sqrt{2\pi \frac{2\bar{D}_i}{\bar{u}_i^2} t^3}}\mathrm{e}^{\left(-\frac{(t - \frac{z_i}{\bar{u}_i})^2}{2\frac{2\bar{D}_i}{\bar{u}_i^2}t}\right)} \nonumber\\
    &=\frac{\mu_i(z_i)}{\sqrt{2\pi\theta_i(z_i) t^3}}\mathrm{e}^{\left( -\frac{(t-\mu_i(z_i))^2}{2\theta_i (z_i) t}\right)} = f_\mathrm{IG}(t,z_i;\mu_i,\theta_i),
    \label{eq:proof-pdf}
\end{align}
which exactly matches the \ac{PDF} of the \ac{IG} distribution in \eqref{eq:pipe-pdf}. This concludes the proof. 
\end{proof}

\subsection{Proof of Theorem 2}
\label{sec:proof2}
\begin{proof}
    Applying the one-sided Laplace transform \ac{wrt} $z_i$ to the path flux $\bar{j}_k(z_i=z_w,t;z_q)$ in \eqref{eq:path-conv} converts the convolutions in the time domain to multiplications in the frequency domain
    \begin{align}\label{eq:MultiplicationsInLaplaceDomain}
        \mathcal{L}\{\bar{j}_k(z_w,t;z_q)\} &= \bar{J}_k(z_w,s;z_q)\nonumber\\ &= J_q(l_q-z_q,s)\cdot J_w(z_w,s)\cdot\hspace*{-3mm}\prod_{i\in\mathcal{E}_k\backslash\{q,w\}} \hspace*{-3mm}J_i(l_i,s)\,.
    \end{align}
    Inserting \eqref{eq:jilap} into \eqref{eq:MultiplicationsInLaplaceDomain} yields 
    \begin{align}
        &\bar{J}_k(z_w,s;z_q) =\nonumber\\ &\mathrm{e}^{\left(\lambda_q(s,\bar{u}_q,\bar{D}_q)(l_q-z_q)+\lambda_w(s,\bar{u}_w,\bar{D}_w)z_w+\sum_{i\in\mathcal{E}_k\backslash\{q,w\}}\lambda_i(s,\bar{u}_i,\bar{D}_i) l_i \right)}\,.
        \label{eq:conv-s}
    \end{align}
    For homogeneous parameters in pipes $p_i\in P_k$, i.e., $\bar{u}_i = \bar{u}$ and $\bar{D}_i = \bar{D}$, all $\lambda$-values in \eqref{eq:conv-s} are equal, i.e., 
    \begin{align}
\lambda_i(s,\bar{u}_i,\bar{D}_i) &= \lambda_q(s,\bar{u}_q,\bar{D}_q)=\lambda_w(s,\bar{u}_w,\bar{D}_w) \nonumber\\ &= \frac{\bar{u} - \sqrt{\bar{u}^2 + 4\bar{D} s}}{2\bar{D}} = \lambda (s,\bar{u},\bar{D}),
    \end{align}
and the path flux in the Laplace domain follows as
    \begin{align}
        \hspace*{-2mm}\bar{J}_k(z_w,s;z_q) &= \mathrm{e}^{\left(\lambda(s,\bar{u},\bar{D})\left(l_q - z_q + z_w + \sum_{i\in\mathcal{E}_k\backslash\{q,w\}} l_i \right)\right)} \nonumber\\ &=\mathrm{e}^{\left(\lambda(s,\bar{u},\bar{D})\bar{l}_k(z_q,z_w)\right)},
        \label{eq:path-flux-s}
    \end{align}
    where $\bar{l}_k(z_q,z_w)$ is the effective path length given in~\eqref{eqn:path_length}. Taking the inverse Laplace transform of \eqref{eq:path-flux-s}, and applying the modifications already used in \eqref{eq:proof-pdf} exactly yields the \ac{PDF} of an \ac{IG} distribution with mean, variance, and scale parameter according to \eqref{eq:pathMom}. This concludes the proof.
\end{proof}

\subsection{Proof of Theorem 3}\label{sec:proof3}
Below, we first derive the cumulants and moments of the true path \ac{FPT}~\eqref{eq:sumt} and the approximate path \ac{FPT} in~\eqref{eq:tk-het}. Second, we show that the error between the skewness of the true and approximate path \ac{FPT} is negligibly small in the considered molecule transport regime. 
For a more compact mathematical description, and without loss of generality, we assume a path $P_k$ with the \ac{Tx} placed at the inlet of the first pipe $p_q$ in the path, i.e., at $z_q=0$. Additionally, we observe the flux at the outlet of the last pipe $p_w$ in the path, i.e., at $z_w=l_w$. In doing so, we can drop the explicit positional arguments $z_q$ and $z_w$ in $p_q$ and $p_w$ for better readability.

\begin{proof}
    First, we consider the \ac{FPT} $T_i$ of a single pipe $p_i$ and derive the \ac{MGF} as 
    \begin{align}
        M_{T_i}(s) &= \mathrm{E}[\mathrm{e}^{sT_i}] = \int_0^\infty \mathrm{e}^{st}f_\mathrm{IG}(t,z_i;\mu_i,\theta_i)\,\mathrm{d}t = \nonumber\\
        &=\mathrm{e}^{\left({\frac{\mu_i}{\theta_i}(1 - \sqrt{1 - 2\theta_is})}\right)},
        \label{eq:momgen}
    \end{align}
    where $\mathrm{E}[\cdot]$ denotes the expectation operator.
    From the \ac{MGF}, the \ac{CGF} can be obtained as 
    \begin{align}
        K_{T_i}(s) = \mathrm{log}M_{T_i}(s) = \frac{\mu_i}{\theta_i}(1 - \sqrt{1 - 2\theta_is})\,.
        \label{eq:cgf}
    \end{align}
    From \eqref{eq:cgf}, the first three cumulants of $T_i$ follow as 
    \begin{align}
        \kappa_1(T_i) &= K_{T_i}'(0) = \mu_i,
        && \kappa_2(T_i) = K_{T_i}''(0) = \mu_i\theta_i,\nonumber\\
        \kappa_3(T_i) &= K_{T_i}'''(0) = 3\mu_i\theta_i^2, 
        \label{eq:cums}
    \end{align}
    where $K_{T_i}'(0)$, $K_{T_i}''(0)$, and $K_{T_i}'''(0)$ denote the first, second, and third derivatives of $K_{T_i}(s)$ \ac{wrt} $s$, evaluated at $s = 0$. Then, from the first three cumulants in \eqref{eq:cums}, the first three moments of $T_i$ are obtained as 
    \begin{align}
        \mathrm{E}[T_i] &= \kappa_1(T_i) = \mu_i, \label{eq:mom-mean}\\
        \mathrm{Var}[T_i] &= \kappa_2(T_i) = \mu_i\theta_i = \sigma_i^2, \label{eq:mom-variance}\\
        \mathrm{Skew}[T_i] &= \xi_i = \frac{\kappa_3(T_i)}{\kappa_2^{\nicefrac{3}{2}}(T_i)} = 3\sqrt{\frac{\theta_i}{\mu_i}},\label{eq:mom-skew}
    \end{align}
    where $\mathrm{Var}[\cdot]$ and $\mathrm{Skew}[\cdot]$ denote the variance and skewness operator, respectively. 
    Eqs.~\eqref{eq:mom-mean} and \eqref{eq:mom-variance} confirm the moments provided in \eqref{eq:pipe-moms} for a single pipe. 
    
    Based on the cumulants and moments of a single pipe $p_i$, below, we first derive the skewness of the true path \ac{FPT} $\bar{T}_k$ in \eqref{eq:sumt}, and subsequently the skewness of the approximated path \ac{FPT} $\hat{T}_k$ in~\eqref{eq:tk-het}.
    As all pipe \acp{FPT} are mutually independent (see Section~\ref{ssec:pathmodel}), $\bar{T}_k$ is the sum of the individual pipe \acp{FPT} $T_i$ (see \eqref{eq:sumt}), and the path PDF is obtained by convolution of the individual pipe PDFs (see \eqref{eq:path-conv}). Hence, for the moment and cumulant generating functions as well as the cumulants, it follows that
    \begin{align}
        M_{\bar{T}_k}(s) &= \prod_{i\in\mathcal{E}_k}M_{T_i}(s), 
        &&K_{\bar{T}_k}(s) = \sum_{i\in\mathcal{E}_k}K_{T_i}(s),\\
        \bar{\kappa}_m(\bar{T}_k) &= \sum_{i\in\mathcal{E}_k}\kappa_m(T_i), &&\forall m \geq 1\,. \label{eq:path-cum3}
    \end{align}
    Then, the skewness of the true path \ac{FPT} $\bar{T}_k$ in \eqref{eq:sumt} can be derived from the definition of the skewness in \eqref{eq:mom-skew} and the path cumulants in \eqref{eq:path-cum3} as 
    \begin{align}
        \mathrm{Skew}[\bar{T}_k] = \bar{\xi}_k &= \frac{\bar{\kappa}_3(\bar{T}_k)}{\bar{\kappa}_2^{\nicefrac{3}{2}} (\bar{T}_k)} = \frac{\sum_{i\in\mathcal{E}_k}\kappa_3(T_i)}{\left(\sum_{i\in\mathcal{E}_k}\kappa_2(T_i)\right)^{\nicefrac{3}{2}}} \nonumber
        \\
        &= 3\frac{\sum_{i\in\mathcal{E}_k}\mu_i\theta^2_i}{\left( \sum_{i\in\mathcal{E}_k}\mu_i\theta_i\right)^{\nicefrac{3}{2}}}\,.
        \label{eq:skew-tru}
    \end{align}

    For the approximated path \ac{FPT} $\hat{T}_k$ of path $P_k$ in \eqref{eq:tk-het}, the skewness $\hat{\xi}_k$ cannot be derived directly from the cumulants, but is implicitly defined by the matched moments in~\eqref{eq:mom-het-mu} and~\eqref{eq:mom-het-sigma}, and the scale parameter in~\eqref{eq:mom-het-theta} as 
    \begin{align}
        \mathrm{Skew}[\hat{T}_k] = \hat{\xi}_k = 3\sqrt{\frac{\hat{\theta}_k}{\hat{\mu}_k}} \overset{\eqref{eq:mom-het-mu}-\eqref{eq:mom-het-theta}}{=} 3\frac{\sqrt{\sum_{i\in\mathcal{E}_k}\mu_i\theta_i}}{\sum_{i\in\mathcal{E}_k}\mu_i}. \label{eq:skew-app}
    \end{align}
    
    In the last part of the proof, we derive an expression for the error $\Delta\xi$ between the skewness of the true \eqref{eq:skew-tru} and the approximated \eqref{eq:skew-app} \acp{FPT} of path $P_k$, i.e., 
    \begin{align}
        \Delta\xi = \bar{\xi}_k - \hat{\xi}_k\,.
        \label{eq:sk-err1}
    \end{align}
    To obtain a tractable expression, we define the relative weights
    \begin{align}
        &w_i := \frac{\mu_i}{\sum_{j\in\mathcal{E}_k}\mu_j} = \frac{\mu_i}{\hat{\mu}_k}\quad \mathrm{with}\quad 
        \sum_{i\in\mathcal{E}_k} w_i= 1,
        \label{eq:weights}
    \end{align}
    where the denominator is equivalent to the matched path mean in \eqref{eq:mom-het-mu}. Further, we define the weighted mean of the $\theta_i$-values~as
    \begin{align}
        m_1 := \sum_{i\in\mathcal{E}_k}w_i\theta_i = \frac{\sum_{i\in\mathcal{E}_k}\mu_i\theta_i}{\hat{\mu}_k} = \hat{\theta}_k,
        \label{eq:meanthetha}
    \end{align}
    which is equivalent to the matched path scale parameter in~\eqref{eq:mom-het-theta}.
    Exploiting this, we can define the weighted variance of the $\theta_i$-values as 
    \begin{align}
        \text{Var}_\mathrm{w}(\theta_i) := \sum_{i\in\mathcal{E}_k} w_i(\theta_i - m_1)^2\,.
        \label{eq:vartheta}
    \end{align}
    Note that the variance in~\eqref{eq:vartheta} is measure for the heterogeneity of the pipes $p_i$ in a path $P_k$.
    Using \eqref{eq:weights}--\eqref{eq:vartheta}, the expressions for the true \eqref{eq:skew-tru} and approximated \eqref{eq:skew-app} path \acp{FPT} can be reformulated as 
    \begin{align}
        \bar{\xi}_k\hspace*{-.75mm} &= \hspace*{-.75mm}3\frac{\sum_{i\in\mathcal{E}_k}\mu_i\theta^2_i}{\left( \sum_{i\in\mathcal{E}_k}\mu_i\theta_i\right)^{\nicefrac{3}{2}}} \hspace*{-.75mm}= \hspace*{-.75mm}3\frac{\bar{\mu}_k\sum_{i\in\mathcal{E}_k}w_i\theta_i^2}{\bar{\mu}_k^{\nicefrac{3}{2}}m_1^{\nicefrac{3}{2}}} \nonumber\hspace*{-.75mm}=\hspace*{-.75mm}\frac{3}{\sqrt{\bar{\mu}_k}}\frac{\sum_{i\in\mathcal{E}_k}w_i\theta_i^2}{m_1^{\nicefrac{3}{2}}},\\
        \hat{\xi}_k &= 3\frac{\sqrt{\sum_{i\in\mathcal{E}_k}\mu_i\theta_i}}{\sum_{i\in\mathcal{E}_k}\mu_i} = 3\frac{\sqrt{\hat{\mu}_k m_1}}{\hat{\mu}_k} = \frac{3}{\sqrt{\hat{\mu}_k}}\sqrt{m_1}\,.
    \end{align}
    Inserting these expressions into \eqref{eq:sk-err1}, and exploiting \eqref{eq:weights}--\eqref{eq:vartheta}, and $\hat{\mu}_k = \bar{\mu}_k$ (see \eqref{eq:mom-het-mu}), it follows that 
    \begin{align}\label{eq:skew_error}
        \Delta\xi &= \bar{\xi}_k - \hat{\xi}_k  \overset{\eqref{eq:weights},\eqref{eq:meanthetha}}{=} \frac{3}{\sqrt{\hat{\mu}_k}}\left[
        \frac{\sum_{i\in\mathcal{E}_k}w_i\theta_i^2}{m_1^{\nicefrac{3}{2}}} - \sqrt{m_1} 
        \right]\nonumber\\
        &= \frac{3}{\sqrt{\hat{\mu}_k}}\frac{\sum_{i\in\mathcal{E}_k}w_i\theta_i^2 - m_1^2}{m_1^{\nicefrac{3}{2}}} \overset{\eqref{eq:vartheta}}{=} \frac{3}{\sqrt{\hat{\mu}_k}}\frac{\text{Var}_\mathrm{w}(\theta_i)}{m_1^{\nicefrac{3}{2}}}\nonumber\\
        &= 3\frac{\text{Var}_\mathrm{w}(\theta_i)}{m_1^{2}}\sqrt{\frac{\hat{\theta}_k}{\hat{\mu}_k}}\,.
    \end{align}
    Finally, by inserting the definition of the path Péclet number $\mathrm{Pe}_k = 2 \hat{\mu}_k/\hat{\theta}_k$ into~\eqref{eq:skew_error} and keeping $\mathrm{Var}_\mathrm{w}(\theta_i)$ fixed, the error $\Delta\xi$ between the true \eqref{eq:skew-tru} and approximated skewness \eqref{eq:skew-app} follows as 
    \begin{align}
        \Delta\xi = \bar{\xi}_k - \hat{\xi}_k = 
        3\frac{\text{Var}_\mathrm{w}(\theta_i)}{m_1^2}\sqrt{\frac{2}{\mathrm{Pe}_k}}\propto \frac{1}{\sqrt{\mathrm{Pe}_k}},
    \end{align}
    which is inversely proportional to the path Péclet number $\mathrm{Pe}_k$. From this, it is evident that the proposed \ac{MIGHT} model is physically accurate for large path Péclet numbers, e.g., as assumed in the Aris-Taylor regime~\cite{Wicke18}.
    This concludes the proof. 
\end{proof}

\bibliography{bibliography_all_authors}

\end{document}